\documentclass[12pt]{article}
\usepackage{fullpage,epsfig,graphics,amsbsy,amssymb,cancel}
\usepackage{psfrag,hyperref}
\usepackage{graphicx,color}
\usepackage{amsfonts}
\usepackage{amssymb,amsmath,amscd}

\newcommand{\BS}{\boldsymbol}
\newcommand{\BB}{\mathbb}

\newcommand{\bea}{\begin{eqnarray}}
\newcommand{\eea}{\end{eqnarray}}
\newcommand{\beq}{\begin{eqnarray}}
\newcommand{\eeq}{\end{eqnarray}}
\newcommand{\nn}{\nonumber}
\newcommand{\Tr}{\textrm{Tr}}
\def\L{{\cal L}}
\def\M{{\cal M}}
\def\N{{\cal N}}
\def\R{{\cal R}}
\newcommand{\smalint}{{\Large\textrm{$\smallint$}}}
\newcommand{\Imb}{\textrm{Imb}}

\begin{document}
\setcounter{page}{0}
\thispagestyle{empty}
\begin{flushright} \small
UUITP-22/10  \\
 \end{flushright}
\smallskip
\begin{center} \LARGE
{\bf Knot Invariants and New Weight Systems from General 3D TFTs}
 \\[12mm] \normalsize
{\large \bf Jian~Qiu and Maxim~Zabzine} \\[8mm]
 {%\small
 \it
   Theoretical Physics, Department of Physics and Astronomy,\\
     Uppsala university,
     Box 516,
     SE-75120 Uppsala,
     Sweden\\}
\end{center}
\vspace{10mm}

\begin{abstract}
\noindent We introduce and study the Wilson loops in a general 3D
topological field theories (TFTs), and show that the expectation
value of Wilson loops also gives knot invariants as in
Chern-Simons theory. We study the TFTs within the
Batalin-Vilkovisky (BV) and
Alexandrov-Kontsevich-Schwarz-Zaboronsky (AKSZ) framework, and the
Ward identities of these theories imply that the expectation value
of the Wilson loop is a pairing of two dual constructions of
(co)cycles of certain extended graph complex (extended from
Kontsevich's graph complex to accommodate the Wilson loop). We
also prove that there is an isomorphism between the same complex
and certain extended Chevalley-Eilenberg complex of Hamiltonian
vector fields. This isomorphism allows us to generalize the Lie
algebra weight system for knots to weight systems associated with
any homological vector field and its representations. As an
example we construct knot invariants using holomorphic vector
bundle over hyperK\"ahler manifolds.
\end{abstract}

%\noindent

\eject \normalsize \eject

\tableofcontents

\section{Introduction}
\label{intro} Knots are embeddings of $S^1$ into some ambient
space which is usually $\mathbb{R}^3,S^3$. It is the global
feature of the knot, namely, how it is embedded that is of the
most interest. One is interested in studying the cohomology of the
space of embeddings, which we denote as Imb. This space is
disconnected, different components are separated by walls
corresponding to singular knots. This leads to the natural
definition of knot invariants as $\tilde{H}^0(\Imb)$ whose
elements are locally constant functions on Imb. The knot
polynomials are classes of this group that behave
multiplicatively under the multiplication of knots. For example,
the well known Jones polynomial \cite{Jones} assigns
\begin{figure}[h]
\begin{center}
\psfrag {n}{$n$}
\includegraphics[width=2in]{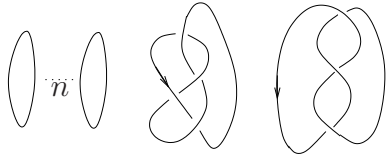}
\caption{$n$ trivial links, figure 8 knot and the braid knot
$s^3$}\label{example_knots_fig}
\end{center}
\end{figure}
$q^{-(n-1)/2}(-q-1)^{n-1}$, $q^{-2}(1-q+q^2-q^3+q^4)$ and $q(1+q^2-q^3)$
to the knots in fig.\ref{example_knots_fig}.

Interestingly a simple device called the \emph{chord diagram} and
its extensions appeared repeatedly in the study of knots. A chord
diagram by definition consists of a circle and bunch of chords
connecting pairs of points of the circle. First of all, if one
denotes by $\Sigma$ the complement of Imb in the space of
$C^{\infty}$ mappings of $S^1$ into $S^3$ or $\BB{R}^3$, one can
gain knowledge about $\tilde H^{\cdot}(\Imb)$ by computing
$H_{\cdot}(\Sigma)$. The chord diagrams made their appearance here
as the labelling of the cells of $\Sigma$. Vassiliev used this
device in his direct computation of the group
$\tilde{H}_{\cdot}(\Sigma)$ \cite{Vass}.

The chord diagram and its extension can also be given the
structure of a differential complex $({\cal G}^{\cdot},\delta)$,
which leads to another independent construction of classes in
$H_{dR}^0(\Imb)$ inspired by Chern-Simons (CS) perturbation
theory, due to Bott and Taubes \cite{BottTaubes} and many others.
They studied the de Rham instead of the singular cohomology of Imb
through integrating certain tautological forms over the
configuration space (this step is known as the \emph{transfer
map}). This construction builds 'models' for classes in
$H_{dR}^0(\Imb)$ and the models are labelled by the cycles of the
extended graph complex (all of these will be reviewed later).
After carefully blowing up the singular points of the
configuration space, which for a physicist means the
regularization of short distance singularity, the de Rham
differential in $\Omega^{\cdot}(\Imb)$ is related through the
transfer map to the differential $\delta$ of the graph complex
\cite{cattaneo-2002-2}. More precisely, there is a homomorphism
\bea ({\cal
G}^{\cdot},\delta)\stackrel{transfer}{\longrightarrow}(\Omega^{\cdot}(\Imb),d)~.\label{transfer}\eea
Hence by feeding cycles in the graph complex to the lhs of
Eq.\ref{transfer}, one obtains knot invariants from the rhs. At
the same time, Kontsevich invented a different configuration space
integral including only the chord diagrams; it is believed that
two constructions give the same answer \cite{Kontsevich:knot}.

The configuration space integral construction is inspired by the
CS perturbation expansion, where the extended graphs are none
other than the Feynman diagrams, and the transfer map is just the
Feynman integral while the blowing up of singular points are some sort
of Cauchy principle value prescription for regulating the
divergences. The CS theory in 3D, like a myriad of other 3D TFT's
can be neatly formulated in the Batalin-Vilkovisky
(BV)-Alexandrov-Kontsevich-Schwarz-Zaboronsky (AKSZ) framework
\cite{Alexandrov:1995kv,Roytenberg:2006qz}: the BV language
handles the gauge fixing problem with ease while the AKSZ
construction throws the geometrical aspect of the theory into a
sharper focus. We shall show in this paper that the graph
differential of Eq.\ref{transfer} naturally arises out of the
(rather simple) BV yoga and the homomorphism between ${\cal
G}^{\cdot}$ and $\Omega^{\cdot}(\Imb)$ is one of the
manifestations of the Ward identity. The BV machinery
is particularly handy in demonstrating the equivalence between the
Kontsevich integral and the integrals in Bott Taubes's
construction.

The direct proof of the homomorphism Eq.\ref{transfer} is in fact
a one-line proof, but we take a detour of first proving the
isomorphism between the said graph complex and certain extension
of the Chevalley-Eilenberg (CE) complex, which is a generalization
of Kontsevich's earlier result \cite{Kontsevich:knot}. Then we
prove the homomorphism from the CE complex to the de Rham complex.
The first isomorphism gives what is known as the \emph{weight
system}: by constructing the cycles in the CE complex, one finds
cycles for the graph complex which can be fed to the transfer map
to produce knot invariants. So this detour allows one to interpret
the expectation value of the Wilson loop in CS theory as the
pairing between two dual constructions of the graph complex. The
similar interpretation of the 3D TFT partition function as such a
pairing was pointed out in ref.\cite{Kontsevich:Feynamn} and
explained in detail in our earlier work \cite{QiuZabzine:2009rf}.
What is more important, this detour makes it clear what objects
can be used as a weight system. We show in this paper that any
representation of a homological vector field or a $Q$-structure
can be used as a weight system. The $Q$-structure by definition is
a deg 1 vector field on a graded manifold (GrMfld) with $Q^2=0$. A
representation of $Q$ is an extension $Q+Q_R$, which acts on some
vector bundle over the GrMfld, such that $(Q+Q_R)^2=0$. Our main
result is that if $Q$ admits a non-trivial Hamiltonian lift, then
we can define a weight system for knot invariants. So far most of
the weight systems come from the Lie algebras, which is just one
special case of the above general $Q$-structure. Our construction
is inspired by the work of Rozansky and Witten
\cite{Rozansky:1996bq} and Sawon \cite{sawon}. The Rozansky-Witten
(RW) weight system was also discussed in ref.\cite{roberts-2001},
but the discussion there involves much more sophisticated machinery.
The necessity of weight systems other than those from the Lie
algebras is called for after the work of Vogel \cite{Vogel}, who
disproved the conjecture that all weight systems come from
semi-simple Lie algebra (the stronger version, which drops the
word semi-simple is also believed to be true).

The paper is organized as follows: we first review the
construction of knot invariants from Chern-Simons perturbation
theory in sec.\ref{KICST}, there we demonstrate a recurring theme
of this paper which is the factorization of the partition function
or the expectation value of a Wilson loop into the pairing of two
dual constructions of graph co(cycle). After giving a 'picturesque
motivation' for general weight systems, we move on to construct 3D
TFT's whose Feynman rules correspond exactly to these weight
systems. For these theories Wilson loops analogous to the CS
theory can be constructed and we propose the definition of
representation of $Q$-structure in sec.\ref{sec_rep_Q}. And we
show that as far as the perturbation expansion is concerned, the
calculation proceeds exactly as that of CS theory, and hence
non-pathological. In sec.\ref{BV_yoga}, we generalize the
definition of the graph complex to incorporate the Wilson loops.
Even though this was already done by Cattaneo et al
\cite{Cattaneo:1996pz}, we show that the graph differential
follows from the Ward identity in the BV formalism and an
isomorphism between the graph complex and certain extended
Chevalley-Eilenberg complex. Sec.\ref{Conf_Space_Integ} serves as
a tribute to the beautiful work of Bott and Taubes and a concrete
example of the our abstract BV manipulation. In sec.\ref{KIQR}, we
perform a low order sample calculation for the weight system of a
$Q$-structure and its representation associated with hyperK\"ahler
manifolds and holomorphic vector bundles. The knot invariant takes
value in $H_{\bar\partial}$ instead of complex number, which is
the main novelty of our result. We point out some loose ends in
sec.\ref{UNPB}, which contains an apologetic review of Vogel's
work as a justification of our considering new weight systems, and
some speculations of the nature of these weight systems that are
intrinsically different from the conventional ones.

\section{Knot Invariants from Chern-Simons Theory}\label{KICST}

In this section we quickly review the construction of knot
invariants and 3-manifold invariants from Chern-Simons theory
 initiated by Witten \cite{Witten:1988hf}, for a nice review see \cite{Labastida:1998ud}.

Let $G$ be a simple Lie group and consider the connection $A$ of a
principle $G$-bundle over some dim 3 manifold $\Sigma_3$. The
Chern-Simons theory is defined by the path integral as
\bea Z=\int {\cal
D}A\exp\Big(\frac{ik}{4\pi}\smalint_{\Sigma_3}\Tr(A\wedge
dA+\frac{2}{3}A\wedge A\wedge A)\Big)~,\nn\eea%
where one integrates over all the gauge equivalence classes of
connections $A$ weighted by the exponential of the Chern-Simons
functional. In this expression, $k\in \BB{Z}$ is the Chern-Simons
level. Since the theory is formulated with only differential forms
over $\Sigma_3$ and the partition function $Z$ is expected to be a
topological invariant of $\Sigma_3$.

In the large $k$ limit, the path integral is done using the
stationary phase approximation. The stationary points, which is
the solution to the equation of motion, are given by the flat
connections. One then breaks up the gauge field into the flat
background connection $A_i$ and the fluctuation $B$: $A=A_i+B$.
$Z$ is the sum of $Z_{A_i}$ where $A_i$ range over all gauge
equivalence classes of flat connection. $Z_{A_i}$ itself is
obtained by integrating over the fluctuation $B$. To integrate
over the fluctuations gauge fixing is needed to ensure we are not
counting gauge equivalent fluctuations. This is commonly done by
imposing the Lorentz gauge (with the help of a metric on
$\Sigma_3$). The gauge fixed action around $A_i$ is
\bea
S_{GF}=S(A_i)+\frac{ik}{4\pi}\Tr\smalint_{\Sigma_3}Bd_iB+\frac{2}{3}B^3+\phi
d^{\dagger}_i B+\bar cd^{\dagger}_i
(d_ic+[B,c])~,\label{CS_action_GF}\eea
where $d_i$ is the gauge covariant derivative with connection
$A_i$. Integrating out $\phi$ would put us on the Lorentz gauge,
while the ghost anti-ghost $c,\bar c$ provides the Fadeev Popov
determinant.

To the lowest order in $1/k$, there is the one loop determinant.
The norm of the determinant is the Ray-Singer torsion at $A_i$
which is a topological invariant. The phase of the determinant is
more subtle: a gravitational Chern-Simons term is needed to remedy
the anomalous transformation. We gloss over this point as it is
not central to the paper.

Beyond the one loop determinant factor, the higher order
perturbation expansion comes from the Feynman diagram calculation.
For this, it is rather expedient to assemble the various fields
into a super field. Introduce an odd coordinate $\theta^a$ that
transforms like 1-form $dx^a$ on $\Sigma_3$. Define a super field
\bea \BS{A}=c+\theta^aB_a+\frac{1}{2}\theta^b\theta^a\tilde
A_{ab}~,\nn\eea
where $\tilde A=d^{\dagger}_i\bar c$. We note that $\bar c$ was a
3-form originally, and the change of variable from $\bar c$ to
$\tilde A$ causes a Jacobian which is offset exactly by the
Jacobian from integrating out $\phi$. The gauge fixing condition
is now neatly summarized by saying that $\BS{A}$
is co-exact w.r.t $d_i$, and the action condenses to become
\bea
S_{GF}=S_{A_i}-\frac{ik}{4\pi}\int\limits_{\Sigma_3}d^3xd^3\theta\Tr\big(
\BS{A}D_i\BS{A}+\frac{2}{3}\BS{A}^3\big)~,\label{CS_action_GF_SF}\eea
where
$D_i\BS{A}^{\alpha}=\theta^a(\partial_a\delta^{\alpha\gamma}+if^{\alpha\beta\gamma}A^{\beta}_i)\BS{A}^{\gamma}$
is the covariant derivative in the super language and
$\BS{A}=\BS{A}^{\alpha}t_{\alpha}$. The perturbation theory can be
carried out using the super Feynman rules \cite{Axelrod:1991vq}.
It is also worth mentioning that the form of the gauge fixed
action in super language is the motivation of the AKSZ
construction of general TFT's.

The perturbation theory at each order of $1/k$ are also
topological invariants of $\Sigma_3$. The invariance can actually
be explained by exploring the relation between Feynman integral
and certain graph complex, which was first defined by Kontsevich
\cite{Kontsevich:Feynamn}. The proper definition of the graph
complex is collected in the sec.\ref{Extended_Graph_Complex}, but
for now suffice it to say that Feynman diagrams are examples of
graphs and that the graph complex is equipped with a differential,
which acts on a graph by shrinking in turn each of its
propagators.

For the perturbation calculation one first expands the cubic
vertex in Eq.\ref{CS_action_GF_SF} down the exponential and
contract all the fields using the propagators. The resulting
Feynman diagrams are of course all tri-valent. Then for a Feynman
diagram $\Gamma$, one integrates the propagators (with Lie algebra data stripped off)
over the positions of the vertices on
$\Sigma_3$; this step assigns each diagram a number
$b_{\Gamma}$\footnote{It has been established that $b_{\Gamma}$ is
finite by Axelrod and Singer \cite{Axelrod:1991vq}}. And at the
same time, since each cubic vertex carries a structure constant
$f^{abc}$ with it, the indices of $f$ will be contracted together
according to the given $\Gamma$, resulting in a number
$c_{\Gamma}$. Kontsevich realized that $b_{\Gamma},c_{\Gamma}$ can
be used to construct
\bea \sum_{\Gamma}b_{\Gamma}\Gamma^*\ \ \textrm{and}\ \ \
\sum_{\Gamma}c_{\Gamma}\Gamma\label{dual_construction}~.\eea
The former is a cocycle in the graph complex which can be shown
using integration by parts; while the latter is a cycle basically
due to the Jacobi identity. The partition function
$Z=\sum_{\Gamma}b_{\Gamma}c_{\Gamma}$ is just the pairing of the
two dual constructions of the graph complex
\bea Z=\langle\sum_{\Gamma}b_{\Gamma}\Gamma^*,
\sum_{\Gamma}c_{\Gamma}\Gamma\rangle=\sum_{\Gamma}b_{\Gamma}c_{\Gamma}~.\nn\eea
The topological invariance can now be explained roughly as: the
change of metric changes each propagator by an exact form and
integration by part of this exact form will cause the differential
to hit neighboring propagators which gives delta functions and
thereby shrinks the propagators one by one. This manipulation is
exactly like the differential of the graph complex, in other
words, the change of metric changes $b_{\Gamma}$ by a coboundary,
hence
\bea
\delta_{g}Z=\langle\sum_{\Gamma}(\delta_{g}b_{\Gamma})\Gamma^*,
\sum_{\Gamma}c_{\Gamma}\Gamma\rangle=\langle\delta(\cdots),
\sum_{\Gamma}c_{\Gamma}\Gamma\rangle=\langle\cdots,\partial
\sum_{\Gamma}c_{\Gamma}\Gamma\rangle=0~,\label{invariance}\eea
where we denote by $\partial$ the differential of the graph
complex, $\delta$ its dual and $\delta_g$ is the variation w.r.t
the metric.

Before moving on to Wilson loops, we mention in passing that for
certain choices of $\Sigma_3$, an exact formula for $Z$ is known.
This was done by first looking at $Z_{S^2\times S^1}$, which is
equal to 1, and through performing surgeries relating
$Z_{S^2\times S^1}$ to other $\Sigma_3$'s such as the 3-sphere or
the lens space. Major effort has been poured into this arena
\cite{Jeffrey,Rozansky:1994ba,Freed:1991wd} making it possible to
obtain some exact results.

More interesting is the case when there are Wilson loops in the
theory. The Wilson loop is given as
\bea W_R=\Tr_R\BB{P}\exp\big({\scriptsize\textrm{$\oint$}} dt
A\big)~,\nn\eea where $\BB{P}$ means the path ordering and the
trace is taken over representation $R$ of the Lie algebra.

Due to the metric independence, the expectation value
\bea \langle W_R\rangle=\int
DA~\Tr_R\BB{P}\exp\big({\scriptsize\textrm{$\oint$}}A\big)e^{S_{CS}}\nn\eea
is invariant under continuous deformation of the Wilson loop and
so characterizes the topological information of how the Wilson
loop is embedded inside $\Sigma_3$, in other words, these are knot
invariants by construction.

The fact that the perturbation theory produces knot invariants can
be likewise analyzed by taking an excursion to the graph complex,
this time slightly generalized. Now one expands both the cubic
vertex in the action and the Wilson loop operator into power
series. We call vertices from the action the internal ones and
those from the Wilson loop peripheral ones, then the propagators
will run among all the vertices. It is customary to include an
oriented loop into the Feynman diagram to signify the Wilson loop. The resulting Feynman diagrams are tri-valent
for the internal vertices and uni-valent for the peripheral ones.
These will be an example of the extended graph complex which we
define in sec.\ref{Extended_Graph_Complex}, for now we just say
that there is a properly generalized differential for such graph
complex as well.

The Feynman integral prescribes that we should integrate over the
entire $\Sigma_3$ the positions of internal vertices, while for
the peripheral ones along the Wilson loop and respecting the
cyclic order. Similar to the situation earlier, this step assigns
a number $b_{\Gamma}$ for every $\Gamma$. The number $c_{\Gamma}$
is also obtained likewise from Lie algebra data. And no surprise
that the cochains and chains defined in Eq.\ref{dual_construction}
will remain cocycles and cycles once we properly modify the
definition of the differential of the extended complex. This time
the deformation of the Wilson loop will change $b_{\Gamma}$ by a
coboundary and invariance of the path integral can be analyzed as
in Eq.\ref{invariance}. Note the $c_{\Gamma}$ for the extended
graphs are called \emph{weight systems} for knots.

The knot polynomials in the introduction are characterized by the
so called \emph{skein relation}. In practice, one calculates the
knot invariant by untying the knot until one reaches the trivial
knot. In the process, one needs to let two strands of the knot
pass each other, and the skein relation dictates the change to the
value of the knot polynomial incurred in each such passing.
Finally, the value of the knot polynomial is the sum of the
cumulative changes. So obtaining the skein relation is absolutely
central. In certain situation the skein relation in the
Chern-Simons theory can also be obtained exactly, also through the
surgery formula. What one does is to gouge out of $\Sigma_3$ a
small ball containing two strands of the Wilson loop and glue back
the ball but twisted by a diffeomorphism of the boundary. In the
few Jones polynomials listed under fig.\ref{example_knots_fig},
the $q$ is just $\exp\big(2\pi i/(k+2)\big)$, where $k$ is the
Chern-Simons level.

\begin{figure}[h]
\begin{center}
\psfrag {a}{\scriptsize{$a$}}\psfrag {b}{\scriptsize{$b$}}\psfrag
{c}{\scriptsize{$c$}}\psfrag {d}{\scriptsize{$d$}}\psfrag
{par}{\scriptsize{$\partial$}}\psfrag{del}{\scriptsize{$\delta$}}
\psfrag{I}{I}\psfrag{H}{H}\psfrag{X}{X}
\includegraphics[width=2.5in]{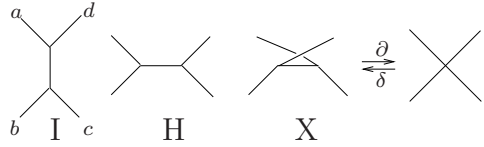}
\caption{Lie algebra weight system}\label{LieAlgWght_fig}
\end{center}
\end{figure}

So far we have seen that we can break apart the partition of
Chern-Simons theory with or without Wilson loops into the two
independent parts of Eq.\ref{dual_construction}, making it easier
to study/generalize them. In what comes next, we will first
generalize the Lie algebra weight system, since it has been an
unresolved problem if all weight system comes from Lie algebras.
We saw from Eq.\ref{invariance} that the only requirement for
$c_{\Gamma}$ is that $\sum_{\Gamma}c_{\Gamma}\Gamma$ should be a
cycle in the graph complex. The key quality of the Lie algebra
weight system that meets this requirement is the Jacobi identity.
Indeed, take three graphs that are identical except in two of the
vertices as depicted in fig.\ref{LieAlgWght_fig}. The Feynman rule
from the Lie algebra weight system assigns
\bea c_s=f^{ade}f^{bce};\ \ c_t=f^{abe}f^{cde};\ \
c_u=f^{bde}f^{cae}\nn\eea to the $s$-,$t$- and $u$-channel
diagram. The graph differential contracts the central propagator
and all three channels collapse into the four point vertex. The
Jacobi identity then says $c_s+c_t+c_u=0$ indicating that this is
a graph cycle. This graph relation is hieroglyphically denoted as
the 'IHX' relation. As a variant to IHX, we can place the lower edge of I and both of the two vertices of H X
on a Wilson loop with a given representation. This time IHX relation simply says the representation 'represents' the Lie algebra and is called the STU relation.

\begin{figure}[h]
\begin{center}
\psfrag {n}{\scriptsize{$n$}}\psfrag
{n-l}{\scriptsize{$n-l$}}\psfrag {l}{\scriptsize{$l$}}\psfrag
{sum}{\scriptsize{$\sum_{l=2}^{n-2}$}}\psfrag
{del}{\scriptsize{$\partial$}}
\includegraphics[width=2.5in]{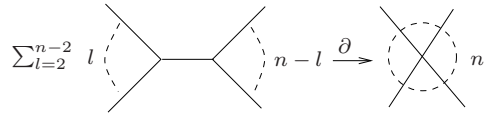}
\caption{$L^{\infty}$ algebra weight system}\label{LinftyWght_fig}
\end{center}
\end{figure}

The general situation when a differential acts on a (not
necessarily trivalent) graph is drawn in fig.\ref{LinftyWght_fig}.
This figure points out clearly that an $L^{\infty}$
algebra\footnote{$L^{\infty}$ structure is most easily formulated
in terms of a homological vector field $Q$, $Q^2=0$. We use these
two terms interchangeably in this paper.} with a hamiltonian lift
will be the natural generalization of Lie algebra weight system in
the following way. The $L^{\infty}$ algebra is a generalization of
Lie algebra in the sense that the Jaccobi identity fails in a
controlled manner. For simplicity, we consider the vector space
with an even symplectic structure $(\BB{R}^{2n|m},\Omega_{AB})$,
take an odd Hamiltonian function $\Theta$ satisfying
$\{\Theta,\Theta\}=0$. Denote by $\Theta_{A_1\cdots A_n}$ the
Taylor coefficient $\partial_{A_1\cdots A_n}\Theta$, and for the
time being assume the Taylor coefficient of $\Theta$ has no linear
or quadratic term. Then up to quartic order the identity
$\{\Theta,\Theta\}=0$ reads
\bea \Theta_{ABE}(\Omega^{-1})^{EF}\Theta_{CDF}+\textrm{perm }
\small{\textrm{$ABCD$}}=0~.\nn\eea
The last relation is the generalization of the Jaccobi identity. We can define the weight
system by assigning the tensor $\Theta_{A_1\cdots A_l}$ and
$\Theta_{A_{l+1}\cdots A_n}$ to the two vertices of
fig.\ref{LinftyWght_fig}. It is imaginable that after
straightening out the sign factors, the rhs of
fig.\ref{LinftyWght_fig} exactly equals
$\partial_{A_1}\cdots\partial_{A_n}\{\Theta,\Theta\}$, i.e. the
rhs of fig.\ref{LinftyWght_fig} is zero and we have graph cycles.
The general case when $\Theta$ does have linear and quadratic
Taylor coefficient corresponds to the 'controlled breach' of
Jaccobi identity mentioned earlier; and we will deal with this in
sec.\ref{Wght_H_Q}.

To conclude, we need to invent a TFT such that the various Taylor
coefficient of $\Theta$ will appear as interaction vertices. The
AKSZ construction for TFT answers this call well.

\section{AKSZ Topological Field Theory}
We need a TFT that can incorporate the data of a Hamiltonian
function satisfying $\{\Theta,\Theta\}=0$. This is systematically
achieved by the AKSZ construction. The reader may see
ref.\cite{QiuZabzine:2009rf} for a more detailed account.

The data for the AKSZ construction is a triple
$(\M,\Omega,\Theta)$, where $\M$ is a graded manifold (GrMfld)
with deg2 symplectic structure $\Omega$. We use $X^A$ as the
coordinate of $\M$. The fields in the theory will be the mappings
\bea \textrm{Maps}(T[1]\Sigma_3,\M)~,\nn\eea
where $\Sigma_3$ is some 3D manifold. For this reason, we call
$T[1]\Sigma_3$ the source manifold and $\M$ the target manifold.
$T[1]\Sigma_3$ is itself a GrMfld, it is the total space of the
tangent bundle of $\Sigma_3$. Denote by $(\xi^a,\theta^a)$ the
coordinates of the base and fibre; $T[1]$ signifies that we assign
$\theta^a$ degree 1. We see $\theta^a$ transforms just like
$d\xi^a$ and with the same commutativity property. Any function
$\BS{F}(\xi,\theta)$ on $T[1]\Sigma_3$ can be expanded in power
series of $\theta$ as
\bea
\BS{F}(\xi,\theta)=F(\xi)+\theta^aF(\xi)_a+\frac{1}{2}\theta^b\theta^aF(\xi)_{ab}
+\frac{1}{3!}\theta^c\theta^b\theta^aF(\xi)_{abc}~.\label{expan_theta}\eea
Each term in the expansion is just a differential form over
$\Sigma_3$, and we call each component $F_{(0)},F_{(1)},F_{(2)}$,
etc. The mapping Maps$(T[1]\Sigma_3,\M)$ will also be described by
the superfields $\BS{X}^A=X^A(\xi,\theta)$ which can be likewise
expanded.

The deg 2 symplectic form $1/2\Omega_{AB}dX^A\wedge dX^B$ induces
naturally a deg $-1$ symplectic form on the mapping space
Maps$(T[1]\Sigma_3, {\cal M})$ according to
\bea \omega=\frac{1}{2}\int\limits_{T[1]\Sigma_3} d^6z~\big
(\Omega_{AB}\delta\BS{X}^A\delta\BS{X}^B\big)~,\label{odd_symp}\eea
where we write $\xi,\theta$ collectively as $z$ and $d^6z\equiv
d^3\theta ~d^3 \xi$. And we can define a Laplacian $\Delta$ such
that $\Delta^2=0$,
\bea \Delta \equiv \int\limits_{\Sigma_3}
d^3\xi~(\Omega^{-1})^{AB}(-1)^{AB}\big(\frac{\delta}{\delta{X}^A_{(3)}(\xi)}\frac{\delta}{\delta{X}^B_{(0)}(\xi)}
+\frac{\delta}{\delta{X}^A_{(1)}(\xi)}\frac{\delta}{\delta{X}^B_{(2)}(\xi)}\big)~.\nn\eea%
The action is
\bea S=S_{kin}+S_{int}=\int\limits_{T[1]\Sigma_3} d^6z~
\frac{1}{2}\BS{X}^A\Omega_{AB} D\BS{X}^B-(-1)^3\BS{\Theta}~,\ \ \
\ D:=\theta^a\partial_a~,\label{oddCSaction}\eea
where $D$ is just the de Rham differential written in super
language. The kinetic term $\BS{X}^A\Omega_{AB} D\BS{X}^B$ is a
sloppy notation, one should really pick up a Liouville form $\Xi$
such that $d\Xi=\Omega$ and write the kinetic term as $\BS{\Xi}$.
When $\partial\Sigma_3=\emptyset$ the dependence on the choice of
$\Xi$ drops once we expand $\BS{\Xi}$ into components, see
ref.\cite{AKSZ_RW}. The sign in front of $\Theta$ is $-(-1)^d$ for
theories on $\Sigma_d$ and it is chosen such that the equation of
motion reads $D\BS{X}=\{\BS{\Theta},\BS{X}\}$.

The action is deceptively simple, the nontrivial part is to define the path integral.
We need to first pick a Lagrangian
submanifold (LagSubMfld) w.r.t $\omega$ of Eq.\ref{odd_symp}. We
recall a submanifold $\L$ of a symplectic manifold $(\N,\omega)$
is Lagrangian iff $\L$ is maximal submanifold such that  $\omega|_{\L}=0$
 (in finite dimensional setting it is middle dimensional submanifold).
For our application here, $\L\in\textrm{Maps}(T[1]\Sigma_3,\M)$
must be chosen such that the symplectic form Eq.\ref{odd_symp}
vanishes when restricted to it. The choice of $\L$ is called
\emph{gauge fixing} since it generalizes the BRST gauge fixing
procedure \cite{Schwarz:1992nx}. The path integral integrates the
super fields $\BS{X}$ over $\L$,
\bea Z_{AKSZ}=\int_{\L}{\cal D}\BS{X}\
\exp\big(-(S_{kin}+S_{int})\big)~.\nn\eea%
The key advantage of this construction is that the complicated
gauge fixing issue is encapsulated in the choice of $\L$, TFT
models like the Rozansky-Witten (RW) model
ref.\cite{Rozansky:1996bq} and RW model coupled to CS
\cite{Kapustin:2009cd} both emerge from very simple AKSZ actions;
these models only assumed their sophisticated form after some
particular choice of $\L$. But most importantly, the various Ward
identities that are responsible for the metric independence and
other invariance of these models are nothing more than the
consequence of the following \bea 0=\int_{\L}{\cal D}\BS{X}\
\Delta(\cdots)~.\nn\eea
Though it is not always easy to find ${\cal L}$, we can nonetheless analyze the general
properties of the path integral by using this equation.

The construction may seem all too abstract, but as we have seen
familiar theories like the CS fit snugly into this framework. For
CS theory, one only need to start from $\M=g[1]$ with $g$ being
some Lie algebra whose coordinate we call $A^{\alpha}$. The
symplectic form is taken as the Killing form of the Lie algebra
$\Omega_{\alpha\beta}=\eta_{\alpha\beta}=\Tr[t_{\alpha}t_{\beta}]$
and $f_{\alpha\beta\gamma}A^{\alpha}A^{\beta}A^{\gamma}$ is used
as $\Theta$. The CS action is the same as Eq.\ref{CS_action_GF_SF}
\bea \int\limits_{T[1]\Sigma_3}d^6z~ \big[\eta_{\alpha\beta}~
\BS{A}^\alpha D\BS{A}^\beta +\frac{i}{3}
f_{\alpha\beta\gamma}\BS{A}^{\alpha}\BS{A}^{\beta}\BS{A}^{\gamma}\big]~,\label{CS_action_AKSZ}\eea
except now the superfield $\BS{A}^{\alpha}$ is
\emph{unconstrained}. And it is not hard to verify that the
constraint $\BS{A}$ being co-exact as in Eq.\ref{CS_action_GF_SF}
corresponds to a choice of LagSubMfld on which the symplectic form
\bea \omega=\frac{1}{2}\int\limits_{T[1]\Sigma_3} d^6z~\big
(\eta_{\alpha\beta}\delta\BS{A}^{\alpha}\delta\BS{A}^{\beta}\big)\nn\eea
vanishes. The geometrical nature of the AKSZ construction allows
one to quickly populate the spectrum of TFT's
\cite{Roytenberg:2006qz,Cattaneo:2001ys,Cattaneo:2009zx,AKSZ_RW,Ikeda:2010vz},
much of the theory remain unexplored, though.

We pause to ask, for an arbitrary $\M$, is the AKSZ theory even a
sensible quantum theory? As far as the path integral formulation
is concerned, there is no difficulty. Since the Gaussian integral
will be understood after proper Wick rotations. The determinant
$\det\Omega$ is nowhere vanishing, so once we have chosen a sign
for $\sqrt{\textrm{det}\Omega}$ there is no more ambiguity.
However to quantize the AKSZ theory is quite a different thing.
The quantization of the two special cases: CS or RW theory is
worked out in ref.\cite{Axelrod:1989xt} and \cite{Rozansky:1996bq}
respectively and is highly non-trivial. Within the scope of this
paper, we can only proceed with the assumption that the general
AKSZ theory is quantum mechanically non-pathological.

The true innovation of the AKSZ theory comes when one looks at the
structure of the partition function. The Feynman rule or the
weight system, $c_{\Gamma}$ assigns every vertex a tensor that is
the Taylor expansion of $\Theta$. The Jacobi identity for Lie
algebra is replaced with the more general relation
$\{\Theta,\Theta\}=0$. This was extensively explored in our
previous paper \cite{QiuZabzine:2009rf}, showing that the path
integral is a handy tool to construct characteristic classes
related to $\Theta$.

Now we need to include the Wilson loop into the general theory and
thereby recruit any such $\Theta$ to act as weight systems for
knots.

The Wilson loop like any other observables in a TFT must be gauge
invariant, which in the BV language implies the following
\bea \delta_B{\cal O}=\{S_{kin}+S_{int},{\cal O}\}=0\ \textrm{and}\ \Delta{\cal
O}=0~,\label{def_BRST}\eea
where $\delta_B$ is the BRST transformation in BV language.
The second condition is \emph{formally} fulfilled for all 3D theories. By using
\bea\{S_{kin},f(\BS{X})\}=-Df(\BS{X})~;~~~~~
\{S_{int},f(\BS{X})\}=\{\Theta,f\}(\BS{X})~,\ \ f\in
C^{\infty}(\M)\nn\eea
the first equation when written in components is
\bea &&0=\{\Theta,{\cal O}\}_{(0)}~,\nn\\%
&& d{\cal O}_{(0)}=\{\Theta,{\cal O}\}_{(1)}~,\nn\\%
&& d{\cal O}_{(1)}=-\{\Theta,{\cal O}\}_{(2)}~,\nn\\%
&&\cdots~,\nn\eea
where the subscript ${}_{(p)}$ means the $p$-form component in the
$\theta$-expansion, see Eq.\ref{expan_theta}. One can solve for
this equation easily if ${\cal O}$ is a line operator insertion by
mimicking the formula for the parallel transport. Pick a curve
$\phi(t),\;t\in[0,1]$ embedded in $\Sigma_3$, and we define the
matrix $U$ as the path ordered exponentiation\footnote{From now on, we adopt the notation that if
any function of the target $\M$ is in boldface, such as $\BS{T}$
in Eq.\ref{wilsonloop}, it is regarded as a function of the
superfield $\BS{X}^A$: $\BS{T}=T(\BS{X})$.}
\bea U(t,0)=\BB{P}\exp\big(-\int_0^tdtd\theta^t\ \BS{T}\big)=\BB{P}\exp\big(-\int_0^tdt\ T_t    \big)~,\label{wilsonloop}\eea
where $T_t$ is a matrix defined as
\bea
(T_t)^A_{B}=\partial_{\theta^t}\BS{T}^A_{\ B}|_{\theta=0} =\partial_C T^A_{tB} (X_{(0)})X^C_{(1)}~.\nn\eea
{\color{black}Strictly speaking we should write in the exponential $\phi^*(T_t)$: the pull back of the 1-form $T_t$ on $\Sigma_3$ to the curve.
The pull back $\phi^*$ will be dropped from now on.}

Assuming that
\beq
\BS{T}^A_{\
\,B}=(\Omega^{-1})^{AC}\partial_C\partial_B\BS{\Theta}~,\nn
\eea
 we can quickly verify the relations
\bea &&\partial_tU^A_{\ B}(t,0)=-T^A_{tC}U^C_{\ B}(t,0),\ \
\partial_tU^A_{\ B}(1,t)=U^A_{\ C}(1,t)T^C_{tB}~,\nn\\%
&&\sum_B(-1)^BT^A_{\ B}T^B_{\ C}+(-1)^A\{\Theta,T^A_{\ C}\}=0~.\label{prop_U}\eea
The second line is a consequence of $\{\Theta,\Theta\}=0$. The
BRST transformation (defined in Eq.\ref{def_BRST}) of $U$ is
{\small\bea \delta_B U^C_{\ D}(1,0)&=&(-1)^{C+A}\int_0^1dt U^C_{\
A}(1,t)\big(\partial_{\theta^t}(-D\BS{T}^A_{\
B}+\{\BS{\Theta},\BS{T}^A_{\ B}\})|_{\theta=0}\big)U^B_{\ D}(t,0)\nn\\%
&=&(-1)^{C+A}\int_0^1dt U^C_{\ A}(1,t)\big(-\partial_tT^A_{\
B}+\partial_{\theta^t}\{\BS{\Theta},\BS{T}^A_{\
B}\}|_{\theta=0}\big)U^B_{\ D}(t,0)~.\nn\eea}%
Integrate by part the
$t$ derivative and use relation Eq.\ref{prop_U}, we get besides a surface term the following
{\small\bea&&(-1)^C\int_0^1dt U^C_{\ E}(1,t)\big((-1)^AT^E_{t
A}T^A_{\ B}-(-1)^ET^E_{\ A}T^A_{tB}+(-1)^E\partial_{\theta^t}\{\BS{\Theta},\BS{T}^E_{\ B}\})|_{\theta=0}\big)U^B_{\ D}(t,0)\nn\\%
&&=(-1)^C\int_0^1dt U^C_{\
E}(1,t)\big(\partial_{\theta^t}((-1)^A\BS{T}^E_{\ A}\BS{T}^A_{\
B}+(-1)^E\{\BS{\Theta},\BS{T}^E_{\
B}\}|_{\theta=0}\big)U^B_{\ D}(t,0)=0~.\nn\eea}%
The surface term is
\bea(-1)^{C+A}U^C_{\ A}(1,t)\big(T^A_{\
B}(t)\big)U^B_{\ D}(t,0)\Big|^1_0=T^C_{\
B}(1)U^B_{\ D}(1,0)-(-1)^{C+A}U^C_{\ A}(1,0)T^A_{\ D}(0)~.\nn\eea
{\color{black}This shows that if the curve is closed, the trace
\bea W=\sum_A(-1)^AU(1,0)^A_{~A}\nn\eea
is BRST invariant and hence a valid observable. The sign factor $(-1)^A$ is reminiscent of the interpretation due to Witten
that the Wilson loop should be regarded as the partition of a quantum mechanics system attached to the curve.}

In the following sections, we will understand the BRST invariance
of the Wilson loop from a Lie algebra chain complex point of view,
there the BRST invariance will also be related to the closeness of
certain graph chain.

\section{$Q$-Structures and Their Representations}\label{sec_rep_Q}
In CS theory, the gauge fields perched on the Wilson loop can be
in representations other than the adjoint one, the same can be
achieved for a general TFT given by the data $(\M,\Omega,\Theta)$.
We need first the notion of a representation.

For a GrMfld $\M$, a deg 1 vector field $Q(X)$ satisfying
$Q^2(X)=0$ is called a homological vector field or a
$Q$-structure. The previous nilpotent Hamiltonian function
$\Theta$ on $(\M,\Omega)$ with $\deg\Omega=2$ induces a
$Q$-structure $Q\cdot=\{\Theta,\cdot\}$. Of course, any
$Q$-structure on $\M$ has a Hamiltonian lift on $T^*[2]\M$
since $T^*[2]\M$ is trivially a symplectic GrMfld with a deg 2
symplectic form. But as we shall see later, $\Theta(X)$ associated
with a trivial lift of some $Q(X)$ tends to give TFT's with
trivial perturbation theory. So it is the 'genuine' nilpotent
Hamiltonian functions that are of interest to us.

For a $Q$-structure over $\M$, consider a vector bundle
$E\rightarrow\M$. {\color{black}We assume the fibre coordinates of $E$ are bosonic even though the generalization to
graded vector bundle is straightforward (compare with Eq.\ref{prop_U}).} Naming the coordinate of the fibre of $E$
as $\xi_a$, we define the representation of a $Q$-structure as an
extension of $Q$ into $Q_R$, with $Q_R^2=0$,
\bea &&Q_R(X)=\{\Theta(X),\cdot\}+{\cal
R}^a_{\ b}(X)\xi_a\frac{\partial}{\partial\xi_b}~,\nn\\
&&0=Q_R^2=\{\Theta,R^a_{\ b}\}\xi_a\partial_{\xi_b}+({\cal R}^a_{\
b}{\cal R}^b_{\ c})\xi_a\partial_{\xi_c}\Rightarrow\{\Theta,{\cal
R}\}+{\cal R}^2=0~,\label{Q_equiv}\eea
Here the matrix ${\cal R}$ may itself be a function of $\M$.

The trivial example is of course the representation of a Lie
algebra. In this case, ${\cal M}$ is $g[1]$: the linear space of
Lie algebra with degree shifted up by 1. We name the coordinate
$A^{\alpha}$. The bundle $E$ is the trivial one $g[1]\times V$
with $V$ being the representation of the Lie algebra. We extend
$Q$ as
\bea
Q=f_{\alpha\beta\gamma}A^{\alpha}A^{\beta}\frac{\partial}{\partial
A^{\gamma}}~;\ \ Q_R=Q+2iA^{\alpha}(t_{\alpha})^i_{\
j}\xi_i\frac{\partial}{\partial \xi_j}~,\label{Lie_Q_R}\eea where
$t_{\alpha}$ is the some matrix satisfying
$[t_a,t_b]=if^c_{ab}t_c$, $\eta_{\alpha\beta}$ is the Killing
metric and
$\eta_{\gamma\delta}f^{\delta}_{\alpha\beta}=f_{\gamma\alpha\beta}$
is totally antisymmetric.

For a general representation we can form a BRST invariant line
insertion the same way as in Eq.\ref{wilsonloop}, it is written as
\bea U(1,0)=\Tr_R\BB{P}\exp\big(-\int_0^1dtd\theta^t\, \BS{{\cal
R}}\big)~,\label{Wilsonloop_rep}\eea
where $\BS{{\cal R}}$ is the matrix ${\cal R}^a_{\ b}(\BS{X})$.
The pull back of 3D superfields (functions on $T[1]\Sigma_3$)
  to 1D superfields (functions on $T[1]S^1$) is again implicitly understood.
{\color{black}The verification of BRST invariance is totally similar and more importantly one can check that under a change of trivialization of $E$, the
Wilson loop changes by a BRST exact piece}\footnote{{\color{black}The authors would like to thank Ezra Getzler for pointing this out for us.}}. In the
next subsection, we give some more examples of $Q$-structure, for
some we can construct interesting representations, for others we
have only the ``adjoint" representation.

\subsection{Examples $Q$-Structure and Representations}\label{EQSR}
As a cousin of the CS theory, there is the BF theory. The relevant
GrMfld for the BF theory is $\M=g[1]\oplus g^*[1]$, where $g^*$ is
the dual of the Lie algebra. We name the coordinate of the two
summands as $A^{\alpha}$ and $B_{\alpha}$. Suppose $V$ is the
representation of the Lie algebra, $E=\M\times (V\oplus V)$, with
$\xi,\eta$ as the coordinates of the two copies of $V$. The $Q$
structure is the same as in Eq.\ref{Lie_Q_R}, the extension is
\bea Q_R=Q+2i(t_{\alpha})^i_{\
j}\big(A^{\alpha}(\xi_j\frac{\partial}{\partial
\xi_j}+\eta_i\frac{\partial}{\partial
\eta_j})+B^{\alpha}\xi_i\frac{\partial}{\partial
\eta_j}\big)\nn\eea
defines a representation of $Q$. In this expression, we have used
the inverse metric $\eta^{\alpha\beta}$ to raise the index on $B$.

The action can also be written down according to the AKSZ
construction
\bea S_{BF}=\int\limits_{T[1]\Sigma_3} d^6z
\BS{B}_{\alpha}D\BS{A}^{\alpha}+\BS{B}_{\alpha}f_{\beta\gamma}^{\alpha}\BS{A}^{\beta}\BS{A}^{\gamma}~.\nn\eea
One can use this representation to form the Wilson loop in BF
theory as in Eq.\ref{Wilsonloop_rep}. Due to the index structure
on $A$ and $B$, any number of $A$'s can be traced together, while
the $B$'s must be placed between two $A$'s. So what one gets is
the 'beaded' Wilson loops. This type of Wilson loops and the knot
invariants were studied in ref.\cite{Cattaneo:1996pz}.

\emph{The canonical representation for a Lie algebroid.} A Lie
algebroid is formulated most easily in the graded manifold
language. The data required is a deg 1 GrMfld $\M$, such a GrMfld
is necessarily of the form $L[1]M$, where $L$ is a vector bundle
over some manifold $M$. We denote by $x^{\mu},\ell^A$ the even and
odd coordinates. Any deg 1 vector field is necessarily of the form
\bea Q=2\ell^AA_A^{\mu}(x) \frac{\partial}{\partial
x^{\mu}}-f^A_{BC}(x)
\ell^B\ell^C\frac{\partial}{\partial\ell^A}~.\nn\eea%
Requiring $Q^2=0$ puts constraint on the coefficients
\bea&&A_{[A}^{\nu}\partial_{\nu}A_{B]}^{\mu}=A^{\mu}_Cf^C_{AB}~,\nn\\
&&A_A^{\mu}\partial_{\mu}f^D_{BC}+f^D_{AX}f^X_{BC}+\textrm{cyclic
in \scriptsize$ABC$}=0~,\label{Lie_Alge_compatible}\eea
$A_A^{\mu}$ is the called anchor and $f^A_{BC}$ is the structure
function. The two conditions constitutes the definition of a Lie
algebroid. The AKSZ theory which potentially calculates the
characteristic classes of a Lie algebroid can be constructed
easily \cite{Cattaneo:2009zx}.

It is interesting to come up with some representation of a Lie
algebroid other than the adjoint one. So far we have only the
canonical representation, which is abelian.  Consider the line
bundle $\wedge^{top}L\otimes \textrm{vol}$, and a section $s$.
$Q_R$ is given by
\bea Q_R=Q+\ell^A(\nabla_{\mu}A^{\mu}_A+f^B_{AB})~,\nn\eea%
 where $\nabla_{\mu}A^{\mu}_A$ denotes the divergence with
  respect to $\textrm{vol}$.
The second term is in fact the modular class of the Lie algebroid
$\theta=\ell^A(\nabla_{\mu}A^{\mu}_A+f^B_{AB})\in L^*$. By
construction, $\theta$ is $Q$-closed. Unfortunately, neither the
adjoint nor the canonical representation for a general Lie
algebroid gives any non-trivial Wilson loops. Note that we are
using the term adjoint representation for Lie algebroid in a loose
sense, the reader may consult ref.\cite{AbadMariusMehta} for a
more formal definition and more extensive discussion of
representations. In general, we expect that some extra structure
such as a metric for the Lie algebroid is required to give
interesting Wilson loops and knot invariants.

The previous $Q$'s are all based on the Lie algebra-ish
constructions, and the TFT constructed from them is Chern-Simons
like. We can also have $Q$-structures on purely even GrMfld's, in
which case we need some odd parameters. As an example, for any
flat vector bundle $E\rightarrow M$ with connection $\Gamma,\
d\Gamma+\Gamma\Gamma=0$, we can form the following GrMfld
$\M=E\oplus T[1]M$. If we name the coordinate of $M$ as $x^{\mu}$,
the fibre of $E$ as $e^i$ and the (odd) fibre of $T[1]M$ as
$v^{\mu}$. The odd coordinate $v^{\mu}$ transforms the same way as
$dx^{\mu}$ and only plays the role of a parameter. The $Q$-structure is
formed as
\bea Q=v^{\mu}\frac{\partial}{\partial x^{\mu}}~;~~~~~~~~~
Q_R=Q+v^{\mu}\Gamma^i_{\mu j}e^j\frac{\partial}{\partial
e^i}~.\nn\eea
$Q_R^2=0$ follows from the flatness condition of $\Gamma$. Once
again, to form TFT's with non-trivial perturbation theory, the
fibre of $E$ must have a symplectic structure. In the Rozansky
Witten model which is built on a hyperK\"ahler manifold, $E$ will
be the holomorphic tangent bundle equipped with the holomorphic
symplectic form. The odd coordinate $v^{\mu}$ may be incorporated
as a dynamic variable or be left simply as a parameter of the
theory. The precise formulae of Rozansky Witten weight system and
its generalization to holomorphic vector bundle will be given in
sec.\ref{Wght_H_Q} and it will be clear there that the knot
invariant takes value in the Dolbeault-cohomology of the base
manifold. Thanks to the work of Beauville and Fujiki
\cite{hyperkahler}, there are now two families of hyperk\"ahler
manifolds to all dimensions, so the weight system from these
families provides substantial generalizations to the Lie algebra
weight systems \cite{roberts-2006,Thompson:2000pw}.

There is a more tantalizing example related to integrable systems which also reaches to all
dimensions. Take $\BB{R}^{2n}$ with the standard even symplectic form
$1/2\Omega_{\mu\nu}d\xi^{\mu}\wedge d\xi^{\nu}$, and suppose we have $n$
functions $f_i$, satisfying $\{f_i,f_j\}=0$. Let us pick $n$ odd
parameters $t^i$ and form the nilpotent Hamiltonian function
$\Theta=\sum_i t^i f_i$, and the AKSZ TFT
\bea S=\int d^6z\
\BS{\xi}^{\mu}\Omega_{\mu\nu}D\BS{\xi}^{\nu}+t^if_i(\BS{\xi})~.\label{integrabmodell}\eea
We do not have a clear notion of a general representation for such
case, but it is likely to be related to the foliation of the phase
space $\BB{R}^{2n}$ by the conserved charges $f_i$. Nor do we have
any clear understanding of the implication of this weight system,
except that for a given integrable system, the perturbation theory
is in fact a finite series due to the anti-commutativity of $t^i$.

\section{Recipe for Calculation}
\label{recipe}
Here we spell out the necessary
detail for the perturbative calculation, even though all the
details were already in ref.\cite{QiuZabzine:2009rf}.

{\color{black}The path integral is defined by the Lagrangian submanifold ${\cal L}$ in the mapping space. The choice of ${\cal L}$ in general
depends on the interaction term $\Theta(X)$. In this section we only focus on the free theory, which means we expand the Wilson loop operator into polynomials of fixed degree and compute its expectation value. Since the theory is free, the equation of motion $D\BS{X}^A=0$ says the saddle point corresponds to closed forms on $\Sigma_3$, in particular the 0-form component $X^A_{(0)}$ are constants. So we are expanding around \emph{a base point on the target space ${\cal M}$}. An alternative point of view is to take the full interacting theory, but evaluate the path integral using background field method around the constant modes.}

From either point of view, the Lorentz gauge used in the CS theory can be applied. To define the Lorentz
gauge, we need to pick a metric $h_{ab}$ for $\Sigma_3$, which
allows us to define the adjoint operator $d^{\dagger}$ for de Rham
differential $d$. As mentioned previously, the super field can be
regarded as poly-forms over $\Sigma_3$, which permits the
decomposition into harmonic (h), exact (e) and co-exact (c) part:
\bea \BS{X}^A=(\BS{X}^A)^h+(\BS{X}^A)^e+(\BS{X}^A)^c~.\nn\eea
The symplectic form is decomposed similarly,
\bea \omega=\frac{1}{2}\int\limits_{T[1]\Sigma_3}
d^6z~\Omega_{AB}\big(
\delta(\BS{X}^A)^h\delta(\BS{X}^B)^h+2\delta(\BS{X}^A)^e\delta(\BS{X}^B)^c\big)~.\nn\eea
One prominent feature is that in this symplectic form, the
harmonic components decouple from the rest. This sector is the
zero modes sector, which can be gauge fixed independently of the
rest. Secondly, the choice $(\BS{X}^A)^e=0$ corresponds clearly to
a valid choice of the LagSubMfld for the nonzero mode sector.
Thirdly, due to the property $\int_{T[1]\Sigma_3}
d^6z~(\BS{X}^A)^c(\BS{X}^B)^c=0$, the quadratic term in the action
drops out, which indirectly excludes all but the tri-valent
vertices in the perturbative expansion (see the counting argument
in sec.5.2 of \cite{QiuZabzine:2009rf}).

\begin{figure}[h]
\begin{center}
\psfrag {R0}{\scriptsize{$-\R^d_{Ce}$}}\psfrag
{R1}{\scriptsize{$-\R^e_{Da}$}}\psfrag {R2}{\scriptsize{$-\R^a_{A
b}$}}\psfrag {R3}{\scriptsize{$-\R^b_{Bc}$}}\psfrag
{R4}{\scriptsize{$-\R^c_{Sd}$}}\psfrag
{T1}{\scriptsize{$-\Theta_{NLF}$}}\psfrag {z}{\scriptsize{$z$}}
\psfrag {z_0}{\scriptsize{$z_0$}}\psfrag
{z_1}{\scriptsize{$z_1$}}\psfrag {z_2}{\scriptsize{$z_2$}} \psfrag
{z_3}{\scriptsize{$z_3$}}\psfrag {0}{\scriptsize{$0$}}\psfrag
{z_4}{\scriptsize{$z_4$}}
\includegraphics[width=1.6in]{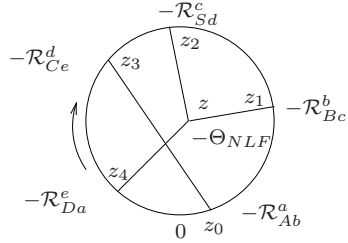}
\caption{A sample diagram, where $R^a_{Ab}=\partial_AR^a_{\
b}(\textsl{x}_0)$, $
\Theta_{ABC}=\partial_A\partial_B\partial_C\Theta(\textsl{x}_0)$}\label{sample_cal_fig}
\end{center}
\end{figure}

As an example, let us take $\Sigma_3$ to be $S^3$  and for the sake of clarity assume that
$\M$ is vector space\footnote{In general curved case when dealing with the perturbation theory
 we will need to apply the exponential map, thus reducing eventually the problem to the linear one.} with even symplectic structure $\Omega$.
 Take odd nilpotent Hamiltonian $\Theta$ and the corresponding representation
$\R$ defined as automorphism of some vector bundle $E$ over $\M$ with the property (\ref{Q_equiv}).
The Taylor coefficients of $\Theta$ and $\R$ will be used as
vertex functions, while $\Omega^{-1}$ appears in conjunction with
the super propagator
\bea \langle
\BS{X}^A(z_1)\BS{X}^B(z_2)\rangle=(\Omega^{-1})^{AB}\BS{G}(z_1,z_2)~.\nn\eea
Take fig.\ref{sample_cal_fig} as an example, for which the time
runs clockwise. The Feynman rules assigns to this figure
{\small\bea &&\left ( \partial_N \partial_L \partial_F
\Theta(\textsl{x}_0) \right )~ \Omega^{NB} \Omega^{LC}
\Omega^{FD}\Omega^{AS}
\Tr_E \big (\partial_A\R(\textsl{x}_0)~ \partial_B\R(\textsl{x}_0) ~\partial_S\R(\textsl{x}_0)~\partial_C \R(\textsl{x}_0)~ \partial_D \R(\textsl{x}_0) \big )\times\nn \\
&&\hspace{2cm}\smalint_{S^3} d^6z\smalint_0^1 d^2z_0
\smalint^{t_0}_0d^2z_1\cdots\smalint^{t_3}_0d^2z_4~\BS{G}(z,z_1)\BS{G}(z,z_2)\BS{G}(z,z_4)\BS{G}(z_0,z_3) \label{sample_cal}\\
&&=c_\Gamma(\textsl{x}_0) \times b_\Gamma~,\hspace{5cm}  \nn \eea}
where we separated the  ``algebraic" factor $c_\Gamma$ and the
kinematic factor $b_\Gamma$. Several points have to be clarified:
The harmonic part of the super field does not participate in the
perturbation theory; for $\Sigma_3=S^3$ we have only the harmonic
0 and 3-form $(X^A_{(0)})^h,(X^A_{(3)})^h$, we let the LagSubMfld
of this sector be given by $(X^A_{(3)})^h=0$, leaving
$(X^A_{(0)})^h$ as a parameter, namely we choose
$(X^A_{(0)})^h=\textsl{x}_0$ as the base point for the Taylor
expansion of $\Theta,\R$. When the target manifold is a vector
space, it may be natural to choose $(X^A_{(0)})^h=0$ as the base
point.
 Consequently, the first line of the formula is
a constant (more generically, the function of zero modes); these are the $c_{\Gamma}$ in the notation of
Eq.\ref{dual_construction}. The second line of the formula gives
$b_{\Gamma}$; it is what inspired the construction of Bott and
Taubes, we discuss this in sec.\ref{Conf_Space_Integ}.

{\color{black}We have discussed the perturbation calculation only in the Lorentz
gauge, but one is free to choose any gauge so long as it is compatible with the separation of zero modes above.}
In fact, Kontsevich's
integral formula for knot invariants \cite{Kontsevich:knot} is
believed to originate from the same CS theory but with light cone
gauge \cite{Labastida:1998ud}. The light cone gauge sets to zero
the cubic vertex in CS action, as a result, there is no internal
vertices in the perturbation theory. On the other hand, a general
gauge fixing may lead to vertices other than trivalent ones.

The strength of BV-AKSZ construction of TFT is that we can discuss
perturbation theory without specifying gauge fixing condition. In
the next section, we derive some important Ward identities for the Wilson loop operator
{\color{black}under the free theory}, which
is true for all gauge fixings {\color{black}with the same zero mode sector}.

\section{Extended Chevalley-Eilenberg and Graph Complex}\label{BV_yoga}
In earlier section, we have touched upon the interplay of Wilson
loops, weight systems and the graph complex. Much of this has been
studied earlier, see
ref.\cite{Cattaneo:1996pz,BN1,Kontsevich:knot} etc. In this
section, we try to understand their relation using the BV
formalism. In the BV framework, the CE differential and graph
differential arise naturally as a consequence of some standard
manipulation, avoiding the untidy analysis used in the discussion
of the metric independence and graph relation in sec.\ref{KICST}.

\subsection{Extended CE Complex}\label{ECE}
We first introduce a Chevalley-Eilenberg (CE) complex taking value
in certain modules. Let $(\BB{R}^{2n|m},\Omega)$ be the super
space with the standard even symplectic structure. We define an
extended Chevalley-Eilenberg chain complex of the Lie algebra of
the Hamiltonian vector fields. The complex is spanned by
\bea
\textrm{CE}_{n,l}=\textrm{span}\big(\BB{X}_{f_0}\wedge\cdots\wedge\BB{X}_{f_n}\otimes
(g_0\otimes\cdots\otimes g_l)\big)\label{CE_chain}~.\eea
And we use the abbreviation
\bea \BB{X}_{f_0}\wedge\cdots\wedge\BB{X}_{f_n}\otimes
(g_0\otimes\cdots\otimes g_l)\stackrel{abbre}{\Rightarrow}
(f_0,\cdots,f_n;g_0,\cdots g_l)\label{abbre}~.\eea
Here $f,g$ are functions over $\mathbb{R}^{2n|m}$ with the
constraint that $f$ is at least quadratic and $g$ is at least
linear (the constraint maybe lifted leading to some interesting
generalizations). $\BB{X}_f$ is the Hamiltonian vector field
induced from $f$. In the second factor cyclic permutation is
allowed (in the graded sense): $(g_0\otimes\cdots\otimes g_l)\sim
(-1)^{(g_0+1)(g_1+\cdots +g_l+l)}(g_1\otimes\cdots\otimes
g_l\otimes g_0)$ and naturally $l+1\equiv0$. Let
$c_{n,l}=(f_0,\cdots,f_n;g_0,\cdots g_l)$, the differential acting
on it is given as\footnote{In the expression for $s_{ij},t_{ij}$
and $u_{nj}$, $f,g$ actually mean $|f|,|g|$, namely the degree of
$f,g$. Similar conventions will appear ubiquitously in the paper.}
\bea
\partial
c_{n,l}&=&\sum_{0\leq i<j\leq
n}(-1)^{s_{ij}}(\{f_i,f_j\},f_0\cdots \hat{f_i}\cdots
\hat{f_j}\cdots f_n; g_0,\cdots,g_l)\nn\\%
&&-\sum_{0\leq i\leq n;0\leq j\leq
l}(-1)^{t_{ij}}(f_0,\cdots\hat{f_i}\cdots f_n;
g_0\cdots g_{j-1},\{f_i,g_j\},g_{j+1}\cdots g_l)\nn\\%
&&-\sum_{0\leq j\leq l}(-1)^{u_{nj}}(f_0,\cdots f_n;
g_jg_{j+1},g_{j+2}\cdots g_l,g_1,\cdots g_{j-1})~,\label{CE_diff}\\
\small{\textrm{$s_{ij}$}}&=&\small{\textrm{$(f_i+1)\sum_{k=0}^{i-1}(f_k+1)+(f_j+1)\sum_{k=0}^{j-1}(f_k+1)+(f_i+1)(f_j+1)+f_i$}}~,\nn\\
\small{\textrm{$t_{ij}$}}&=&\small{\textrm{$(f_i+1)(\sum_{k=i+1}^n(f_k+1)+\sum_{k=0}^{j-1}(g_k+1))+\sum_{k=0}^n(f_k+1)
+\sum_{k=0}^{j-1}(g_k+1)$}}~,\nn\\
\small{\textrm{$u_{nj}$}}&=&\small{\textrm{$\sum_{k=0}^n(f_k+1)+(g_j+1)+\sum_{k=0}^{j-1}(g_k+1)\sum_{m=j}^l(g_m+1)$}}~.\nn\eea

One certainly recognizes the first line as the conventional
differential of the CE complex of Hamiltonian vector fields. The
last part resembles the Hoschild differential and is reminiscent
of the standard bar-resolution of the Lie group, except that ours
has the cyclic property. The second term is simply the action of
the Lie algebra on the second factor of the chain.

The definition of the differential is not wanton, it arises out of
one single Ward identity in the BV formalism, which was first
realized by Schwarz \cite{Schwarz:1999vn},
\bea \int_{\cal L}\Delta(\cdots)=0~.\label{Ward1}\eea
We first list a few useful identities (see \cite{QiuZabzine:2009rf} for more details)
\bea \Delta(\smalint d^{2d}z f(\BS{X}(z)))&=&0~,\nn\\
\Delta\big(\smalint d^{2d}z_1\ f(\BS X(z_1))\smalint d^{2d}z_2\
g(\BS X(z_2))\big)&=&(-1)^{f}\smalint d^{2d}z~\{f(\BS
X(z)),g(\BS X(z))\}~,\nn\\
\big\{\smalint d^{2d}z\ f(\BS X(z)),\smalint d^{2d}z\ g(\BS
X(z))\big\}&=&(-1)^d\smalint d^{2d}z\ \{f(\BS X(z)),g(\BS
X(z))\}~,\nn\\
\big\{S_{kin},f(\BS
X(z))\big\}&=&(-1)^dDf(\BS{X}(z))~,\label{useful}\eea
where $f,g\in C^{\infty}(\M)$ and $d=3$ is the dimension of the
source manifold. One can define a cochain $c_{\L}^{n,0}$ which,
when evaluated on a chain $(f_0,\cdots f_n)$, is given by the path
integral
\bea c_{\cal L}^{n,0}\circ(f_0,\cdots f_n)=\int_{{\cal L}}\smalint
d^6z_0 \BS{f}(z_0)\cdots \smalint d^6z_n
\BS{f}(z_n)e^{-S_{kin}}~.\nn\eea
 Although  we study the formal properties of this path integral expression,
   it makes perfect sense within the perturbative theory.
Applying identities Eq.\ref{Ward1},\ref{useful} one can show that
the cochain is in fact a cocycle w.r.t the differential in
Eq.\ref{CE_diff}:
\bea (\delta c_{\L}^{n,0})\circ(f_0,\cdots f_{n+1})=
c_{\L}^{n,0}\circ\partial(f_0,\cdots f_{n+1})=0~,\ \ \forall\
f_i\in C^{\infty}(\M)~.\nn\eea
The cocycle depends on the choice of the Lagrangian submanifold
${\cal L}$, changing ${\cal L}$ alters the cocycle by a coboundary
\bea (c_{{\cal L}+\delta{\cal L}}^{n,0}-c_{{\cal
L}}^{n,0})\circ(f_0,\cdots f_n)=\tilde c^{n-1,0}\circ\partial
(f_0,\cdots f_n)~,\nn\eea
for some $\tilde c$, the detail is in
ref.\cite{QiuZabzine:2009rf}. We usually drop the subscript
${}_{\cal L}$ on $c_{\cal L}$.

Now we generalize to the case $c^{n,l},\ l>0$ by picking a loop in
$\Sigma_3$ and strewing some $g_i$'s on it in the prescribed
cyclic order.

Assume that the Wilson loop\footnote{Strictly speaking, the line
insertion we are dealing with here is different from that of
Eq.\ref{wilsonloop}, but we use the term Wilson loop nonetheless.}
${\cal K}$ is embedded in $\Sigma_3$ by the function
$\phi^a(t),~t\in [0,1],~a=1,2,3$. We want to describe the
integration of any form over the Wilson loop (a 1-cycle) in the
super language. Take a 1-form $\psi_ad\xi^a$ on $\Sigma_3$ and
rewrite it as a function of $T[1]\Sigma_3$:
$\BS{\psi}(\xi,\theta)=\psi_a\theta^a$, then tautologically we
have
\bea \int_{\cal K}\phi^*\psi=\int_{\cal K}
dt\dot{\phi}^a\partial_{\theta^a}\BS{\psi}=\int_{T[1]{\cal K}}
dtd\theta^t\,\BS{\psi}~.\nn\eea In the last step we have used the
somewhat imprecise notation $d\theta^t$ to denote
$\dot{\phi}^a\partial_{\theta^a}$.   Obviously there is well-defined operation of
 pulling back superfields from $T[1]\Sigma_3$ to $T[1]S^1$.
%
%Let us assume that we have trivialized the tangent bundle of the
%source manifold restricted to the Wilson loop, thereby furnishing
%us with coordinates in the neighborhood of the Wilson loop:
%$z^{\perp}\equiv x^{\perp},\theta^{\perp}$ are the ones transverse
%to the loop; $z^{\|}\equiv  t,\theta^t$ are the ones
%parameterizing the Wilson loop. We also use the Poincare dual of
%the Wilson loop $\BS{u}$ to write the line integral in
%superfield form. Let the $t\in[0,1]$ with $t\sim t+1$ understood, denote%
%\bea W^l(g_0,g_1,\cdots
%g_l)=\int_0^1d^6z_0u(z_0)\BS{g}_0\int_{t_0-1}^{t_0}d^6z_1u(z_1)\BS{g}_1
%\int_{t_0-1}^{t_1}d^6z_2u(z_2)\BS{g}_2\cdots\int_{t_0-1}^{t_{l-1}}d^6z_lu(z_l)\BS{g}_l.\nn\eea
We place $l+1$ insertions $g_i$ on the loop with prescribed cyclic
ordering; this may be written as
{\small\bea W^l(g_0,g_1,\cdots
g_l)=\int_0^1dt_0d\theta^t_0\,\BS{g}_0(z_0)\int_{t_0-1}^{t_0}dt_1d\theta^t_1\,\BS{g}_1(z_1)
\int_{t_0-1}^{t_1}dt_2d\theta^t_2\BS{g}_2(z_2)\cdots\int_{t_0-1}^{t_{l-1}}dt_ld\theta^t_l\BS{g}_l(z_l)~,\label{W}\eea}
where $z$ denotes both $t,\theta^t$. The lower limit of the
integrals is perhaps not what one is used to having in a Wilson
loop. But these integration limits indeed describes $l+1$
insertions that are distributed on the loop with fixed cyclic
ordering. When all the insertions are the same, which is the case
of a true Wilson loop, we can rewrite the lower limit in the
conventional way
\bea W^l(g,g,\cdots
g)=(l+1)\int_0^1dt_0d\theta^t_0\,\BS{g}(z_0)\int_0^{t_0}dt_1d\theta^t_1\,\BS{g}(z_1)
\cdots\int_0^{t_{l-1}}dt_ld\theta^t_l\BS{g}(z_l)~.\nn\eea
We define the cochain $ c^{n,l}$ by the path integral
with both bulk insertions and Wilson loop
\bea c^{n,l}\circ(f_0,\cdots,f_n;g_0,\cdots,g_l)=\int_{{\cal
L}}\smalint d^6z_0 \BS{f}_0\cdots \smalint d^6z_n \BS{f}_n \cdot
W^l(\BS{g}_0,\cdots \BS{g}_l)\ e^{-S_{kin}}~.\label{zero_chain}\eea
And we investigate the Ward identity
\bea 0&=&\int_{{\cal L}}\Delta\big(\cdots e^{-S_{kin}}\big)
=\int_{{\cal L}}\Delta\big(\smalint d^6z_0 \BS{f}_0\cdots \smalint
d^6z_n \BS{f}_n\big)\cdot W^l(\BS{g}_0,\cdots \BS{g}_l)\ e^{-S_{kin}}\nn\\%
&&+(-1)^{\sum_{k=0}^n(f_k+3)}\int_{{\cal L}}\big\{\smalint d^6z_0
\BS{f}_0\cdots \smalint d^6z_n \BS{f}_n, W^l(\BS{g}_0,\cdots
\BS{g}_l)\big\}\ e^{-S_{kin}}\nn\\%
&&-(-1)^{\sum_{k=0}^n(f_k+3)}\int_{{\cal L}}\smalint d^6z_0
\BS{f}_0\cdots \smalint d^6z_n \BS{f}_n\cdot\big\{S_{kin},
W^l(\BS{g}_0,\cdots \BS{g}_l)\big\}\
e^{-S_{kin}}~,\label{Ward_id}\eea
where we dropped terms like $\{S_{kin},\int d^6z\BS{f}\}=\int
d^6zD\BS{f}=0$. $\Delta$ acting on the first part gives a series
of terms resembling the first line of Eq.\ref{CE_diff}, which is
the usual Lie algebra differential \cite{QiuZabzine:2009rf}. We
denote three terms in Eq.\ref{Ward_id} as
$\partial_I,\partial_V,\partial_H$ after ref.\cite{BottTaubes}.
The internal $\partial_I$ collapses two bulk insertions, and the
vertical $\partial_V$ collapses a bulk insertion with one on the
Wilson loop because the effect of the Poisson bracket is that it
removes one $f_i$ and replace one $g_j$ with $\{f_i,g_j\}$. To see
the effect of $\partial_H$, we need to calculate the bracket
$\{S_{kin},W\}$.

The bracket formally gives total derivatives, as
\bea \{S_{kin}, \int_a^b dtd\theta^t
\BS{g}(z)\}=+\int_a^bdtd\theta^t D\BS{g}(z)=\int_a^bdt
\partial_tg~,\nn\eea
but due to the upper and lower limits of the $t$ integral in $W$,
one gets some important surface terms (we set $t_{-1}:=1$ in the
following)
{\small\bea \{S_{kin},W\}
&=&\sum_{k=2}^l(-1)^{\sum_{k=0}^{k-1}(g_m+1)}\int_0^1dt_0d\theta^t_0\;
\BS{g}_0(z_0)\int_{t_0-1}^{t_0}\cdots\nn\\%
&&\int_{t_0-1}^{t_{k-2}}dt_{k-1}d\theta^t_{k-1}\
\BS{g}_{k-1}\BS{g}_k(z_{k-1})
\int_{t_0-1}^{t_{k-1}}\cdots\int_{t_0}^{t_{l-1}}
dt_ld\theta^t_l\;\BS{g}_l(z_l)\nn\\%
&&+(-1)^{(g_l+1)\sum_0^{l}(g_m+1)}\int_0^1dt_0d\theta^t_0\;
\BS{g}_l\BS{g}_0(z_0)\int_{t_0-1}^{t_0}\cdots\int_{t_0}^{t_{l-1}}
dt_ld\theta^t_l\;\BS{g}_l(z_l)~.\nn\eea}%
Written concisely,
\bea \{S_{kin},W\}&=&(-1)^{g_0+1}W(g_0g_1,\cdots,g_l)+\cdots+
(-1)^{\sum_0^{l-1}(g_m+1)}W(g_0,\cdots,g_{l-1}g_l)\nn\\%
&&+(-1)^{(g_l+1)\sum_0^l(g_m+1)}W(g_lg_0,g_1,\cdots,g_{l-1})~,\nn\eea
in other words, the horizontal $\partial_H$ collapses two
neighboring points on the Wilson loop.

Collecting these results, the Ward identity is summarized as
\bea 0&=&(\delta c^{n,l})\circ(f_0,\cdots f_n;g_0,\cdots g_l)\nn\\
&=&\sum_{0\leq i<j\leq n}(-1)^{s_{ij}}
c^{n-1,l}\circ(\{f_i,f_j\},f_0\cdots\hat{f_i}\cdots\hat{f_j}\cdots f_n;g_0,\cdots g_l)\nn\\
&&-\sum_{0\leq i\leq n;0\leq j\leq
l}(-1)^{t_{ij}}c^{n-1,l}\circ(f_0,\cdots,\hat{f_i},\cdots
f_n;g_0,\cdots ,g_{j-1},\{f_i,g_j\},g_{j+1},\cdots g_l)\nn\\
&&-\sum_{0\leq j\leq l}(-1)^{u_{nj}}c^{n,l-1}\circ(f_0,\cdots
f_n;g_jg_{j+1},g_{j+2}\cdots g_l,g_0,\cdots g_{j-1})~,\nn\eea
with the sign factors none other than those of Eq.\ref{CE_diff}.
This tells us that the path integral with both bulk and Wilson
line insertions is a cocycle of the extended CE complex
\bea c^{n,l}\circ\partial((f_0,\cdots,f_n;g_0,\cdots g_l))=(\delta
c^{n,l})\circ(f_0,\cdots,f_n;g_0,\cdots g_l)=0~.\nn\eea
Similarly, $c^{n,l}$ depends on $\L$ according to $c^{n,l}_{\delta
\L}=\delta \tilde c$ for some $\tilde c$. Hence the cohomology
class $[c^{n,l}_{\L}]$ is independent of the continuous
deformation of ${\cal L}$.

\subsection{Extended Graph Complex}\label{Extended_Graph_Complex}

The extended CE complex of the last section is intimately related
to the extended graph complex to be introduced presently.

First recall the definition of an ordinary graph complex. A graph
is a 1-dimensional CW complex (whose vertices need not be
trivalent). We mostly consider closed graphs (without external
legs), the open graphs appear briefly in our review of Vogel's
construction. An important ingredient of the graph is its
orientation. The orientation is given by the ordering of all the
vertices and orienting of all the edges. We remark here that there
are other orientation schemes that are equivalent to the current
one. For example, the orienting of all the edges can be replaced
by the ordering of all the legs from all vertices. Another
convenient scheme is to order the incident legs for each vertex
and order all the even valent vertices. We refer the reader to
sec.2.3.1 of ref.\cite{ConantVogtmann} for a full discussion of
the orientation.

We say the two orientations are the same (resp. opposite) if the
sum of the number of permutations of the vertices plus the number
of flips of edges required to take one orientation to another is
even (resp.odd). Consider the linear space spanned by all the
graphs with all the orientations, then the graph complex is the
quotient of this space by the relation  $\sim:\;
(\Gamma,-or)=-(\Gamma,or)$. Denote by $[\Gamma_{or}]$ the
corresponding element of $(\Gamma,or)$ in the quotient space. A
direct consequence of this quotient is the absence of edges
starting and ending on the same vertex.

The graph complex is equipped with a differential $\partial_I$,
which acts on the graph by collapsing one edge in turn. To be
precise, if an edge runs from the $i$th vertex to the $j$th with
$i<j$, then one collapses the two vertices making a new vertex
labelled by $i$ inheriting all the legs from vertex $i$ and $j$
except the collapsed one. The labelling of vertices after $j$ move
forward one notch. Finally one includes a sign $(-1)^{j+1}$
($(-1)^{j}$ if the collapsed edge runs from $j$ to $i$) for the
resulting graph.

There is an important isomorphism between this graph complex and
the CE chain complex. Let Ham${}_{2n|m}$ be the Lie algebra of
formal Hamiltonian vector fields on $\BB{R}^{2n|m}$, and as usual
we will think of this in terms of the Hamiltonian functions that
generate these vector fields. Let $\textrm{Ham}^0_{2n|m}\subset
\textrm{Ham}_{2n|m}$ be those Hamiltonian functions having zero
constant and linear term in their Taylor expansion and $osp_{2n|m}\subset
\textrm{Ham}_{2n|m}$ be the ones with only quadratic term. The
chain complex $c_{\cdot}(\textrm{Ham}^0_{2n|m},osp_{2n|m})$ is the
$osp_{2n|m}$ co-invariants of usual CE complex of
$\textrm{Ham}^0_{2n|m}$ (the orbits of the $osp$ action). There is
the following isomorphism
\bea {\cal
G}_{\cdot}\sim\lim_{n\rightarrow\infty}CE_{\cdot}(\textrm{Ham}^0_{2n|m},osp_{2n|m})~.
\label{iso_Lie_graph}\eea
The necessity of the direct limit $n\to \infty$ will become clear
shortly.

The extended graph also includes an oriented circle in addition to
the ingredients of the graph complex introduced above. We
distinguish between the internal vertices and vertices on the
circle (peripheral vertices). The edges of the extended graph may
now run amongst both types of vertices. The orientation of such
graphs is determined by the ordering of the internal vertices, the
orientation of the edges and the ordering of the vertices on the
circle.

\begin{figure}[h]
\begin{center}
%\psfrag {v1}{\small{$v_1$}}\psfrag {v2}{\small{$v_2$}}\psfrag
%{u3}{\small{$u_3$}}\psfrag {u1}{\small{$u_1$}}\psfrag
%{u2}{\small{$u_2$}}
\psfrag {v1}{}\psfrag {v2}{}\psfrag {u3}{}\psfrag {u1}{}\psfrag
{u2}{}%
\psfrag {dH}{\scriptsize{$\partial_H$}}\psfrag
{dI}{\scriptsize{$\partial_I$}}\psfrag
{dV}{\scriptsize{$\partial_V$}}
\includegraphics[width=0.9in]{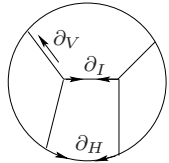}
\caption{Differential of a graph}\label{dec_graph_diff_fig}
\end{center}
\end{figure}

\begin{figure}[h]
\begin{center}
\psfrag {i}{\scriptsize{$i$}}\psfrag {j}{\scriptsize{$j$}}\psfrag
{k}{\scriptsize{$j$}}\psfrag{l}{\scriptsize{$j+1$}}\psfrag
{c}{\small{$(-1)^{j+1}$}}\psfrag {d}{\small{$(-1)^{i+1}$}}\psfrag
{e}{\small{$(-1)^{I+j}$}}
\includegraphics[width=1.8in]{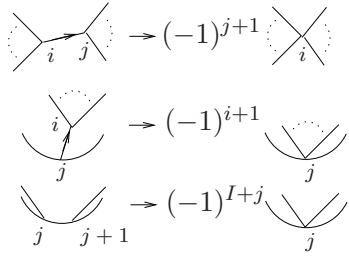}
\caption{Sign factor for differential, here $I$ is the number of
internal vertices. In the third picture, if $j$ is the
$l^{th}$--the last--vertex, then we rename the new vertex 0 with
sign factor $(-1)^{I+l}$.}\label{sign_fac_fig}
\end{center}
\end{figure}

The extended graph also has a differential, which consists of,
apart from $\partial_I$, the collapsing of a pair of adjacent
vertices on the circle $\partial_H$ and the collapsing of one
internal vertex with one on the circle ($\partial_V$).
Fig.\ref{dec_graph_diff_fig} is an illustration of
$\partial_{I,H,V}$.
The sign factors are given in fig.\ref{sign_fac_fig}.
We prove in the appendix that this extended graph complex is
isomorphic to the extended CE complex in a purely algebraic way.
The importance of this isomorphism is: all the properties the
extended CE complex has are enjoyed by the extended graph complex.
Our proof in sec.\ref{ECE} of the Feynman integral as a cocycle of
the extended CE complex can be grafted over and we have the key
result: \emph{the Feynman integral is a cocycle in the graph
complex}.

In fact thanks to this isomorphism, we can prove some stronger
results about the relation between the extended graph complex and
the de Rham complex $\Omega^{\cdot}$(Imb) of the space of
embeddings: ${\cal K}\hookrightarrow\Sigma_3$. Our result is
stronger in the sense that it is valid for any metric of
$\Sigma_3$.

In sec.\ref{ECE}, we took the embedding of the Wilson loop into
$\Sigma_3$ as $\phi^a(t)$, but now we think of it as a function
\bea \textrm{Imb}\times S^1\stackrel{\phi(t)}{\rightarrow}
\Sigma_3~,\nn\eea and write $\phi^a_{\cal K}(t)$ when necessary to
emphasize its dependence on Imb. Thus the Feynman integral, being
a cocycle, evaluates any graph and returns a number, which
actually is a 0-form of Imb. How do we differentiate this
function? In general, to deform a Wilson loop infinitesimally, we
pick a vector field $v$ defined on a neighborhood of the Wilson
loop, and deform the Wilson loop along $v$: $\phi^a\to\phi^a+v^a$.
Let $\psi$ be a 1-form on $\Sigma_3$ that is integrated along the
Wilson loop, we have the following standard manipulation
\bea
\delta_v\int_a^bdt\dot{\phi}^a\psi_a(\phi(t))&=&\int_a^bdt\big(\dot{v}^a\psi_a+\dot{\phi}^a(\partial_b\psi_a)v^b\big)
=\int_a^bdt\big(\partial_t(v^a\psi_a)+\dot{\phi}^b(\partial_{[a}\psi_{b]}v^a)\big)~,\nn\\%
&=&\int_a^b\phi^*\big(d\iota_v\psi+\iota_vd\psi\big)=\int_a^b\phi^*L_v\psi~.\nn\eea
Looking at the definition of Wilson loop insertions Eq.\ref{W},
where the $\BS{g}$'s correspond to the 1-form $\psi$ here, we see
that $\psi$ depends on $\phi$ implicitly through its dependence on
the fields $\BS{X}$. To deform the Wilson loop in this case, we
need only find operator that deforms $\BS{X}$ to
$\BS{X}+L_v\BS{X}$.

This is done easily in the BV formalism. Define
\bea
\Psi_v
%&=&\frac{1}{2}\sum_{p=0}^{3}\sum_{A,B}\frac{1}{p!(3-p)!}e^{a_1\cdots
%a_3}\int d^3x\nn\\%
%&&\Big\{ X^A_{a_1\cdots
%a_p}\Omega_{AB}\big[(v^b\partial_b)X^B_{a_{p+1}\cdots
%a_3}+(\partial_{a_{p+1}}v^b)X^B_{ba_{p+2}\cdots
%a_3}(3-p)\big](-1)^{(3-p)A+B}\Big\}\nn\\
%
&=&\frac{1}{2}\sum_{A,B}\int
d^6z~\BS{X}^A\Omega_{AB}(L_v\BS{X}^B)(-1)^B~,\nn\eea
where $L_v$ is the Lie derivative along $v$. The Cartan formula
gives $L_v=\{d,\iota_v\}$, which may be written in super language
as $\{D,v^a\partial_{\theta^a}\}$. $\Psi_v$ obviously has the
desired property
\bea
\{\Psi_v,\cdot\}&=&\frac{1}{(3-p)!}\sum_{p=0}^3\sum_C(L_vX^C_{b_{p+1}\cdots
b_3})\frac{\partial}{\partial X^C_{b_{p+1}\cdots
b_3}}=\sum_C(L_v\BS{X}^C)\frac{\partial}{\partial
\BS{X}^C}~.\nn\eea
$\Psi_v$ has zero bracket with any bulk insertions $\{\Psi_v,\int
d^6z \BS{f}\}=\int d^6z L_v\BS{f}=\int d^6z D\iota_v\BS{f}=0$ and
its bracket with the line insertions effectively deforms the
Wilson line. But we stress that $\{\Psi_v,{\cal
O}(\BS{X})\}=L_v{\cal O}(\BS{X})$ only when ${\cal O}$ depends on
$\Sigma_3$ only through $\BS{X}$, and for the same reason we
have the counter intuitive relation,
\bea \{\Psi_u,\Psi_v\}=\Psi_{[v,u]}=(-1)\Psi_{[u,v]}.\nn\eea

We apply the Ward identity again,
{\small\bea&&0=\int_{\cal
L}\Delta\Big(\Psi_v\smalint d^6z \BS{f_0}\cdots\smalint d^6z \BS{f_n}\cdot W^l(\BS{g}_0,\cdots \BS{g}_l)\ e^{-S_{kin}}\Big)\Rightarrow\nn\\%
&&\int_{\cal L}\{\Psi_v,\smalint d^6z \BS{f_0}\cdots\smalint d^6z
\BS{f_n}\cdot W^l(\BS{g}_0,\cdots \BS{g}_l)e^{-S_{kin}}\}\nn\\
&&=-\int_{\cal
L}\Psi_v\Delta\Big(\smalint d^6z \BS{f_0}\cdots\smalint d^6z \BS{f_n}\cdot W^l(\BS{g}_0,\cdots \BS{g}_l)e^{-S_{kin}}\Big)~.\nn\eea}
This relation can be rewritten as
\bea &&\delta_v(c^{n,l}\circ(f,\cdots;g,\cdots))=\tilde{c}\circ\partial(f,\cdots;g,\cdots)~,\label{homomorphism}\\
\textrm{where}&& \tilde{c}\circ(f,\cdots;g,\cdots)=-\int_{\cal
L}\Psi_v(\smalint\BS{f}\cdots W(\BS{g},\cdots))e^{-S_{kin}}~.\nn
\eea
From Eq.\ref{homomorphism} we have proved that a Feynman integral
will produce constant functions on Imb if we feed into it cycles
in the extended CE complex. Furthermore by the isomorphism between
graph complex and CE complex, we conclude that the Feynman
integral sends graph cycles to $H^0$(Imb). This is all we need as
far as knot invariant is concerned. But we can forge on and tap
more into the Ward identity.

Let us now regard the vector field $v$ not only as a vector field
on $\Sigma_3$, but also as a vector field on Imb and think of
$\tilde c$ as the contraction of $v$ with a 1-form
$\Omega^1$(Imb). Continuing in this track, we can define $q$-forms
$c^{n,l}_{(q)}\circ(f_0,\cdots f_n;g_0,\cdots
g_l)\in\Omega^q$(Imb). This form, when evaluated on $v_1,\cdots
v_q$ is defined as
\bea
\iota_{v_1}\cdots\iota_{v_q}(c_{(q)}\circ(f,\cdots;g,\cdots))=(-1)^{\frac{q(q-1)}{2}}\int_{\cal
L}\Psi_{v_1}\cdots\Psi_{v_q} (\smalint\BS{f}\cdots\cdot
W(\BS{g}\cdots))e^{-S_{kin}}~.\label{import}\eea
We claim that $c_{(q)}$ maps the extended CE complex
homomorphically to $\Omega^{\cdot}$(Imb). To see this, take $q=2$
as an example, from the Ward identity
\bea 0=\int_{\cal L}\Delta\Big(\Psi_{v_1}\Psi_{v_2}
(\smalint\BS{f}\cdots\cdot
W(\BS{g}\cdots))e^{-S_{kin}}\Big)~,\nn\eea
we get
\bea \int_{\cal L}\ \Psi_{v_1}\Psi_{v_2}
\Delta\big((\smalint\BS{f}\cdots
W(\BS{g}\cdots))e^{-S_{kin}}\big)+\int_{\cal L}\
\Psi_{v_1}\{\Psi_{v_2},(\smalint\BS{f}\cdots
W(\BS{g}\cdots))e^{-S_{kin}}\}\nn\\
-\int_{\cal L}\ \Psi_{v_2}\{\Psi_{v_1},(\smalint\BS{f}\cdots
W(\BS{g}\cdots))e^{-S_{kin}}\}-\int_{\cal L}\
\{\Psi_{v_1},\Psi_{v_2}\}(\smalint\BS{f}\cdots
W(\BS{g}\cdots))e^{-S_{kin}}=0~.\nn\eea
To identify the second and third term, we observe
\bea &&L_u\int_{\L}\Psi_v{\cal
O}(\BS{X})=\int_{\L}\{\Psi_u,\Psi_v{\cal
O}(\BS{X})\}+\int_{\L}\Psi_{[u,v]}{\cal
O}(\BS{X})\nn\\
&&=\int_{\L}\{\Psi_u,\Psi_v\}{\cal
O}(\BS{X})+\int_{\L}\Psi_v\{\Psi_u,{\cal
O}(\BS{X})\}+\int_{\L}\Psi_{[u,v]}{\cal
O}(\BS{X})=\int_{\L}\Psi_v\{\Psi_u,{\cal O}(\BS{X})\}.\nn\eea
So the Ward identity can be rewritten as
\bea \big(-\iota_{v_1}\iota_{v_2}\delta c_{(2)}
+L_{v_2}\iota_{v_1}c_{(1)}-L_{v_1}\iota_{v_2}c_{(1)}-\iota_{[v_2,v_1]}c_{(1)}\big)\circ(f,\cdots;g,\cdots)=0~.\nn\eea
Note the first $\delta$ in the left bracket is the CE
differential, while the last three terms in this bracket is just
the definition of the exterior differentiation. The verification
for general $q$ proceeds similarly.

In summary, we defined a graph cochain $c^{n,l}_{(q)}$ taking
values in $\Omega^{q}$(Imb),
\bea c^{n,l}_{(q)}: {\cal G}_{n,l}\longrightarrow
\Omega^q(\Imb)\label{keyresult}~,\eea
where ${\cal G}_{n,l}$ is the graph  with $n+1$ internal
and $l+1$ peripheral vertices. $c^{n,l}_{(q)}$ has the property
\bea \delta c^p_{(0)}=0~,\ \ dc^p_{(0)}=\delta c^{p-1}_{(1)}~,\
\cdots\ dc^{p}_{(q)}=\delta c^{p-1}_{(q+1)}~,\nn\eea
where we define $c_{(q)}^p= \sum_{n+l=p} c_{(q)}^{n,l}$, $\delta$
is the graph cochain differential and $d$ de Rham differential on
Imb \footnote{One can say that this mapping gives a homomorphism
$({\cal G}^{n,l,s},\delta)\to (\Omega^{3n+l+4-2s}(\Imb),d)$, with
$s$ being the number of edges  (propagators). This is essentially
Eq.\ref{transfer}. Here ${\cal G}_{n,l,s}$ is
  the graph  with $n+1$ internal, $l+1$ peripheral vertices and $s$-edges. ${\cal G}^{n,l,s}$ is its dual object.}.
The first of these properties implies that the knot invariant
constructed by evaluating $c_{(0)}^p$ on a graph cycle depends
only on the class of this cycle in $H_{\cdot}({\cal G})$. The
second one of the identities shows that $c_{(0)}$ when evaluated at cycles of the CE complex is invariant under the deformation of the
knot.

To give a fair assessment of the proof given above, the BV path
integral proof, compact as it maybe, is subject to the many
weaknesses inherent in the path integral approach. Some of the
weakness maybe improved, others not. Just to point out a few, the
statement $\Delta\smalint d^6z \BS{f}=0$ requires regularization.
In odd dimension, the heat kernel regulation of $\Delta$ was used
in ref.\cite{QiuZabzine:2009rf} to cure this. Secondly, any time we
use the fact $\int_{\cal L}\cdots(\smalint d^6z D\BS{f})\cdots=0$,
we need to check the validity of integration by part, since
$\BS{f}$ may collide with other insertions and the singularity
invalidates integration by parts. The Ward identity $\int_{\cal
L}\Delta(\cdots)=0$ is rigorous for finite dimension integral, but
how is it faring in an infinite dimensional setup is beyond our
grip. Despite all these problems the BV formal considerations
capture amazingly well the intricate properties of the
perturbation theory. The purpose of the next section is to review
Bott Taubes construction. This construction is just a special case
of the more general discussion of the current section, with a flat
metric on $\Sigma_3$, and a specific choice of ${\cal L}$.

\subsection{Configuration Space Integral}\label{Conf_Space_Integ}
In this section, we try to make the previous abstract discussion
more concrete, and most importantly, to investigate when does our
path integral proof break down. We do so by reviewing the
Bott-Taubes construction \cite{BottTaubes} and fit their formulae
into our framework.

Bott and Taubes exclusively worked with source manifold $S^3$,
which is represented as $\BB{R}^3$ with an added point
$\{\infty\}$ and a flat metric. They tried to construct ansatz for
closed forms on $\Omega^{\cdot}$(Imb) by mimicking the Feynman
integral. Given a knot parameterized by $\phi^a(t)$,
$\phi(0)=\phi(1)$, consider the following trivial bundle structure
over Imb
\bea \textrm{Imb}\times C^0_{n,l}\subset\textrm{Imb}\times
(S^3)^n\times (S^1)^l~,\nn\eea%
where $S^1$ is identified with its image in $S^3$ under $\phi$ and
the superscript 0 means none of the points on $(S^3)^n\times
(S^1)^l$ are allowed to coincide and no point in $\BB{R}^3$ can be
$\infty$. $C^0_{n,l}$ is called the configuration space, the first
$n$ copies of $S^3$ label the position of the internal vertex and
the $l$ copies of $S^1$ labels the position of the peripheral
vertices. Take any graph in our extended complex, one can write
down a form on $\Imb\times C_{n,l}^0$ as follows: suppose
$C_{n,l}^0$ is parameterized as $(x_1,\cdots,x_n,t_1,\cdots t_l)$,
then for each edge in the graph running from vertex $i$ to $j$,
include a propagator
\bea
\omega_{ij}=\frac{1}{4\pi|x_i-x_j|^3}\epsilon_{abc}(x_i-x_j)^a\wedge
d(x_i-x_j)^b\wedge d(x_i-x_j)^c~,\nn\eea%
where $x_i=\phi(t_{i-n})$ if $i>n$. The form on $\Imb\times
C_{n,l}^0$ we want to construct is the product of all such
$\omega_{ij}$'s; it is called the tautological form in
ref.\cite{BottTaubes}. We stress that $\omega_{ij}$ is a form on
$\Imb\times C_{n,l}^0$, and it can have 'legs' in Imb if $i$ or
$j>n$, because one should write
\bea d\phi^a(t)=\dot{\phi}(t)dt+\delta\phi^a(t)~,\nn\eea and the
variation $\delta\phi$ is a 1-form with leg in Imb.

There is a convenient way of viewing the propagator, let
\bea
G_{ij}:=\frac{\vec{x_i}-\vec{x_j}}{|\vec{x_i}-\vec{x_j}|}~,\label{prop_S}\eea
be a map from $C^0_{2,0}$, $C^0_{1,1}$ or $C^0_{0,2}$ to $S^2$
depending whether $i,j$ both less than $n$, one less one greater
than $n$ or both greater than $n$. $\omega_{ij}$ can be viewed as
the pull back of the volume form $\mu$ on $S^2$
\bea &&\omega_{ij}=G_{ij}^*\mu~,\label{prop_BT}\\
&&\mu=\frac{1}{4\pi r^3}\epsilon_{abc}r^adr^b\wedge dr^c~;\ \ \
\int_{S^2}\mu=1~.\nn\eea
We pause to make the remark that this propagator is exactly the
super propagator in our AKSZ TFT. In the gauge choice scheme of
ref.\cite{QiuZabzine:2009rf}, ${\cal L}$ corresponds to setting
$\BS{X}^e$, the exact part of $\BS{X}$, to zero. The super
propagator is
\bea\langle\BS{X}(u,\theta_1),\BS{X}(v,\theta_2)\rangle&=&
\frac{1}{2}\theta_1^b\theta_1^a\langle
X_{ab}(u),X(v)\rangle-\theta_1^a\theta_2^b\langle
X_a(u),X_b(v)\rangle+\frac{1}{2}\theta_2^b\theta_2^a\langle
X(u),X_{ab}(v)\rangle\nn\\
&=&\frac{1}{2}(\theta_1^b-\theta_2^b)(\theta_1^a-\theta_2^a)\epsilon_{ab}^{\
\ c}\partial_cG(u,v)~,\nn\eea
where $G$ is the same as in Eq.\ref{prop_S}. This is exactly the
propagator given in Eq.\ref{prop_BT}.

The importance of this point is that: since the volume form of
$S^2$ is closed, the propagators will be closed forms in
Imb$\times C_{n,l}^0$ since they are the pull back of closed
forms. Furthermore as $\mu^2=0$, graphs with two edges running
between the same pair of vertices will be set to zero. Thirdly,
the volume form of $S^2$ is always well-defined; so long as we can
extend the mapping $G_{ij}$ from $C_{2,0}^0,C^0_{1,1},C^0_{0,2}$
smoothly to some \emph{compactified} configuration space
$C_{2,0},C_{1,1},C_{0,2}$, the propagator will be well defined.
$G_{ij}$ is most easily extended to $C_{0,2}$: when $t_i=t_j=t$
one simply defines $G_{ij}=\dot{\vec{\phi}}/|\vec{\phi}|$. Since
$\phi$ is an embedding, $\dot{\phi}\neq0$. The extension to
$C_{2,0}$ or $C_{1,1}$ will need some resolution of singularity.
In general, one needs to extend all the $G_{ij}$ defined on
$C_{n,l}^0$ smoothly to some compactified $C_{n,l}$.

After the compactification, the trivial bundle structure becomes
Imb$\times C_{n,l}$ with a \emph{compact} fibre. And the
tautological form is closed on the total space of the bundle. The
instinctive step to take will be to integrate this form fibrewise.
 This step is called the transfer map in
ref.\cite{BottTaubes} and from the way the tautological form is
constructed, the transfer is seen to be a map from the graph
complex to $\Omega^{\cdot}$(Imb). Since $C_{n,l}$ is compact, this
integration is \emph{finite} to start with. One asks is the image
of the transfer map a closed form on Imb. If the answer is yes,
then it will be a knot invariant.

Since the tautological forms are closed, we can apply the Stokes theorem as follows
\bea 0=\int_{C_{n,l}}d\textrm{ taut
forms}=\delta\int_{C_{n,l}}\textrm{taut forms}+\int_{\partial
C_{n,l}}\textrm{taut forms}~.\label{stokes}\eea
The integral of tautological forms over configuration space is
nothing but $b_{\Gamma}$ in Eq.\ref{sample_cal}. Had the fibre
$C_{n,l}$ been a closed manifold, we could have concluded that the
transfer map does give a closed form on Imb. But $C_{n,l}$ has
complicated boundaries in general. We remind the reader that
amongst the summands of tautological forms, there are terms with
one leg in $C_{n,l}$ and one leg in Imb, which has the correct
degree to be integrated over the $\partial C_{n,l}$ leaving a
1-form on Imb.

The boundary comes from various sources. The obvious one is when
two insertions on the Wilson loop collide, which is certainly of
codimension 1 in the configuration space and we have seen how to
compactify these already. The configuration when two internal
points collide appears to be of codimension 3, but since the limit
$\lim_{y\rightarrow x}G_{ij}(x,y)$ depends on from which direction
$y$ approaches $x$, one needs to replace the colliding point with
an $S^2$ to keep track of how the limit is taken. To be more
specific, let $\vec{r}\in S^2$ be the direction of the collision,
and $z$ be the colliding point. Substitute $x=z-\vec{r}\epsilon,
y=z+\vec{r}\epsilon$. Define the limiting value of
$G_{ij}(x,y)|_{x=y}$ as
$\lim_{\epsilon\to0}G_{ij}(x,y)=\vec{r}/|r|\in S^2$. Due to this $S^2$,
the codimension of this singular configuration is $1=3-2$. This is
essentially the blow up. Similarly, the configuration when an
internal point collides with a peripheral point is also a codim 1
boundary. The three singular configurations corresponds neatly to
three types of boundary operations of
fig.\ref{dec_graph_diff_fig}. These three are called the
\emph{principle faces} in ref.\cite{BottTaubes}. It is interesting
that, even though out of the three principal faces, only one comes
naturally, the BV manipulation of the previous section can detect
all three types without any strain.

\begin{figure}[h]
\begin{center}
\psfrag {1}{\scriptsize{$1$}}\psfrag {2}{\scriptsize{$2$}}\psfrag
{3}{\scriptsize{$3$}}\psfrag{4}{\scriptsize{$4$}}
\includegraphics[bb=0 0 78 71,width=.6in]{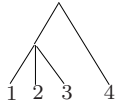}
\caption{Strata of boundary labelled by a grove}\label{grove_fig}
\end{center}
\end{figure}

But how about when $n\geq3$ points colliding? In general, the
stratum structure of the compactified configuration space of $n$
points is labelled by a \emph{grove}. For example $\{\{1,2,3\}\}$
means points 1,2,3 are colliding with a different terminal speed,
and this case also gives a codim 1 boundary. While
$\{\{1,2,3,4\},\{1,2,3\}\}$ means that points $1\sim4$ are
colliding, but to order $\epsilon$, $1,2,3$ are inseparable. Once
we zoom in, we find that 1,2,3 can be resolved by looking up to
order $\epsilon^2$. In this case, the grove looks like
fig.\ref{grove_fig}.
In general each successive blowup implies a closer look at the
singularity and the grove can have many layers. The appendix of
ref.\cite{BottTaubes} and Thurston's thesis \cite{Thurston} give a
very down-to-the-earth review of this.

The codimension of a boundary labelled by a grove is the number of
subsets, so any number of points colliding at a different speed is
also of codim 1 after the blow up. But thanks to the work of Bott
and Taubes, most of these hidden faces do not contribute. The only
worrisome case is when all the points in a graph collide, which
has to be examined case by case. Any how, excluding this last
subtlety, we have demonstrated (up to sign) that the Stokes
theorem Eq.\ref{stokes} can be read as follows (where
$\delta_{{\cal K}}$ is the deformation of knot)
\bea \delta_{\cal K} b_{\Gamma}=b_{\partial\Gamma}\nn\eea
and hence the transfer map is a homomorphism between the graph
complex and $\Omega^{\cdot}(\Imb)$. The same homomorphism for
knots embedded in $\BB{R}^n,\ n>3$ can also be proved
\cite{Cattaneo:1996pz}.

\section{Knot Invariant from $Q$-structure and its Representation}\label{KIQR}
So far we have managed to convert the problem of knot invariants
to the seeking of cycles of the extended graph complex, and the
seeking of weight systems to the cycles in the extended CE
complex. From now on, we solely discuss the problem at the level
of CE complex and forget about the path integral or Chern-Simons
theory.

Out of the three parts the CE differential is made of,
$\partial_I$ is a differential only involving the first part of
the CE chain, and is nilpotent by itself. We know that cycles
w.r.t to $\partial_I$ can be constructed from the $Q$-structures
as in ref.\cite{QiuZabzine:2009rf}. Namely (see the notation of
Eq.\ref{abbre})
\bea c_{n,0}=(\Theta,\cdots\Theta)\nn\eea
is $\partial_I$-closed and simply serves as a weight system for
the 3-manifold invariants. The idea behind is roughly the
  construction for secondary characteristic classes, for
the evaluation of a cocycle on the above cycle is analogous to
plugging in the flat connection of a $G$-bundle to a cocycle of
the CE complex of the Lie algebra of $G$ (see sec.3 of
\cite{QiuZabzine:2009rf}).

The shape of $\partial_V$ suggests that it described some action
of $Q$ on the line insertions $W(g_0,\cdots g_l)$ compatible with
the action of $\partial_H$ which multiplies two adjacent $g$'s. It
is fairly clear that, the object we are trying to sniff out is
some analogue of representation of the $Q$-structure as in
sec.\ref{sec_rep_Q}. We can take a second look of the content of
that section, without the path integral to distract us.

The representation of a $Q$-structure defined in Eq.\ref{Q_equiv},
which we repeat here for convenience
\bea &&Q_R=\{\Theta,\cdot\}+{\cal R}^a_{\ b}
\xi_a\frac{\partial}{\partial\xi_b}~,\nn\\
&&0=Q_R^2=\{\Theta,\R\}+\R^2=0~.\nn\eea
We can immediately form the cycle (in the notation of
Eq.\ref{abbre})
\bea c=\sum_{n+l=N}\frac{1}{n!l}
(\underbrace{\Theta,\cdots,\Theta}_n;\underbrace{\R^{a_1}_{\
a_2},\R^{a_2}_{\ a_3},\cdots,\R^{a_l}_{\ a_1}}_l)~.\label{cycle_Q_R}\eea%
We mention the following to ward off some likely confusions. The
representation matrix $\R^a_{\ b}$ is merely the coefficient of
the chain rather than being a part of the definition of the chain.
To be precise, let us denote by $\{A\},\{B\}$ the collective
indices $\{A_1,\cdots,A_m\}$, $\{B_1,\cdots B_n\}$, and choose the
monomials $x^{\{A\}}=x^{A_1}\cdots x^{A_m}$ as the basis for the
Hamiltonian functions on $\M$. We Taylor expand the $\R$'s and
write the second part of $c$ as
\bea  (\R^{a_1}_{\ a_2},\R^{a_2}_{\ a_3},\cdots,\R^{a_l}_{\
a_1})&=&\pm(\partial_{\{C_1\}}\R^{a_1}_{\
b_1})(\partial_{\{C_2\}}\R^{a_2}_{\ a_3})\cdots
(\partial_{\{C_l\}}\R^{a_l}_{\
a_1})(x^{\{C_1\}},x^{\{C_2\}},\cdots,x^{\{C_l\}})\nn\\%
&=&\pm\Tr\big(\partial_{\{C_1\}}\R\cdot\partial_{\{C_2\}}\R\cdots
\partial_{\{C_l\}}\R\big)(x^{\{C_1\}},x^{\{C_2\}},\cdots,x^{\{C_l\}})~,\nn\eea
where the $\pm$ sign come from pulling the matrices out of the
chain. So only the second factor is the CE chain, while the trace
factor are merely the coefficients. Omitting the indices on $\R$,
the differential acting on $c$ gives
\bea
\partial c&=&\sum_{n+l=N}\frac{1}{n!l}\Big(
-\frac{n(n-1)}{2}(\{\Theta,\Theta\},\underbrace{\Theta,\cdots,\Theta}_{n-2};
\underbrace{\R,\cdots \R}_l)\nn\\%
&&-nl(\underbrace{\Theta,\cdots,\Theta}_{n-1};\{\Theta,\R\},
\underbrace{\R,\cdots
\R}_{l-1})-l(\underbrace{\Theta,\cdots,\Theta}_n;\R\R,
\underbrace{\R,\cdots \R}_{l-2})\Big)\nn\\%
&=&\sum_{n+l=N}\Big(
-\frac{1}{2(n-2)!l}(\{\Theta,\Theta\},\underbrace{\Theta,\cdots,\Theta}_{n-2};
\underbrace{\R,\cdots \R}_l)\nn\\%
&&-\frac{1}{(n-1)!}(\underbrace{\Theta,\cdots,\Theta}_{n-1};\{\Theta,\R\}+\R\R,
\underbrace{\R,\cdots \R}_{l-1})=0~.\nn\eea%

In general one can allow $\xi$ above to have different statistics,
for example when one uses the  ``adjoint" representation. We recall in
that case
\bea \R^A_{\ B}=(\Omega^{-1})^{AC}\partial_C\partial_B\Theta;\
\ \ \sum_C\R^A_{\ C}\R^C_{\ B}(-1)^C+\{\Theta,\R^A_{\ B}\}(-1)^A=0~.\nn\eea%
The cycle is formed in just the same manner, but we warn the
reader to be extra careful with the sign factors in
Eq.\ref{CE_diff} when checking the cycle condition.

\subsection{Weight System Valued in $H_Q$}\label{Wght_H_Q}
In the discussion of the isomorphism between the graph complex and
the CE complex, we required $\Theta$ to be at least quadratic and
$\R$ linear. This restriction of course does not make sense on a
curved manifold, and even when the manifold is flat such as
$\BB{R}^{2n|m}$, the restriction is still too awkward.

The solution in fact brings about a pleasant generalization. We
pick a base point $\textsl{x}_0$ on ${\cal M}$, identify the
neighborhood of $\textsl{x}_0$ with the tangent bundle of ${\cal
M}$ at $\textsl{x}_0$. This is usually done with the exponential
map, but here we only do the simple minded Taylor expansion: the
neighboring points of $\textsl{x}_0$ are parameterized as
$\textsl{x}_0+\xi$ and Taylor expand $\Theta(\textsl{x}_0+\xi),
\R(\textsl{x}_0+\xi)$ into formal power series. The $\xi$ space is
equipped with the symplectic form $\Omega(\textsl{x}_0)$. Denote
\bea
\Theta'=\sum_{n=2}^{\infty}\frac{1}{n!}\partial_{C_1}\cdots\partial_{C_n}\Theta(\textsl{x}_0)\xi^{C_1}\cdots\xi^{C_n}~,\nn\\
\R'=\sum_{n=1}^{\infty}\frac{1}{n!}\partial_{C_1}\cdots\partial_{C_n}\R(\textsl{x}_0)\xi^{C_1}\cdots\xi^{C_n}~,\nn\eea
where the summation starts from 2 and 1 respectively. By Taylor
expanding the equation $\{\Theta,\Theta\}=0$ and
$\{\Theta,\R\}+\R\R=0$ into power series in $\xi$ one clearly sees
that
\bea
&&\{\Theta',\Theta'\}_{\xi}=-2Q^A(\textsl{x}_0)\frac{\partial}{\partial\textsl{x}^A_0}\Theta'~,\nn\\%
&&\{\Theta',\R'\}_{\xi}+\{\R',\R'\}_{\xi}=-Q^A(\textsl{x}_0)\frac{\partial}{\partial\textsl{x}^A_0}\R'-\R\R'-\R'\R~,\label{Q_equiv_gen}\eea
where $Q^A(\textsl{x}_0)$ is $-(\Omega^{-1})^{AB}\partial_B\Theta$
evaluated at $\textsl{x}_0$. In these equations we have written
$\{,\}_{\xi}$ to stress the bracket is taken in $\xi$ space. These
two equations imply $\Theta',\R'$ satisfy the same relation as
Eq.\ref{Q_equiv} \emph{up to $Q(\textsl{x}_0)$-exact term}.

We use $\Theta'$ and $\R'$ to form the following chain
\bea
c(\textsl{x}_0)=\sum_{n+l=N}\frac{1}{n!l}(\underbrace{\Theta',\cdots,\Theta'}_n;\underbrace{\R'^{a_1}_{\
a_2},\R'^{a_2}_{\ a_3},\cdots,\R'^{a_l}_{\ a_1}}_l)~,\nn\eea%
and it depends now on $\textsl{x}_0$. Suppressing the indices on
$\R$ henceforth and using Eq.\ref{Q_equiv_gen} plus the graded
cyclicity we have
\bea
\partial c(\textsl{x}_0)
&=&Q(\textsl{x}_0)\cdot\Big(\sum_{n+l=N}
\frac{1}{(n-1)!l}(\underbrace{\Theta',\cdots,\Theta'}_{n-1};
\underbrace{R',\cdots R'}_l)\Big)~.\nn\eea
Seeing $c(\textsl{x}_0)$ is closed up to $Q$-exact terms, the
proper generalization is clearly staked out: we should talk about
cycles in the CE complex (or graph complex) with coefficients in
the $Q$-cohomology group $H_Q$. This change is easily incorporated
in the path integral. We split any field into
$X^A=\textsl{x}_0^A+\xi^A$, where $\textsl{x}_0$ is the zero mode,
and the path integral is only over the non-zero modes. The details
are in ref.\cite{QiuZabzine:2009rf}.  Thus in practice $c(\textsl{x}_0)$ can
 be calculated by using the perturbation theory, as for example using the Lorentz gauge
  as in Section \ref{recipe}.

This concept of weight system valued in $H_Q$ was already in
ref.\cite{Rozansky:1996bq} and further developed in
ref.\cite{sawon,kapranov-1997}. The construction in
ref.\cite{Rozansky:1996bq} is that on a hyperK\"ahler manifold
with a holomorphic symplectic form $\Omega_{ij}$,
%we have the Bianchi identity
%$\bar\partial_{[\bar i}R_{\bar j]i\ k}^{\ \ \;j}=0$
one can define power series in $\xi^{i}$
\bea\Theta_{\bar
i}=\sum_{n=0}^{\infty}\frac{1}{(n+3)!}\xi^{i_1}\cdots\xi^{i_{n+3}}(\nabla_{i_1}\cdots\nabla_{i_n}R_{\bar
ii_{n+1}\ i_{n+3}}^{\ \ \ \ \ j})\Omega_{ji_{n+2}}~,\nn\eea%
where $R_{\bar ii\ l}^{\ \ k}$ is the curvature tensor and
$\Theta,R,\Omega$ are also implicitly functions of
$\textsl{x}_0^{\bar i}$, which plays the role of the base point
above. $\Theta$ satisfies
\bea\bar\partial_{[\bar j}\Theta_{\bar
i]}=-\frac{1}{2}\{\Theta_{[\bar j},\Theta_{\bar
i]}\}_{\xi}~.\nn\eea%
Introducing a formal deg1 parameter $v^{\bar i}$, which behaves
like $d\textsl{x}_0^{\bar i}$, we can define a $Q$-structure
\bea Q_{rw}=v^{\bar i}\partial_{\bar i}+\{v^{\bar i}\Theta_{\bar
i},\cdot\}_{\xi}~.\nn\eea
Since $\Theta_{\bar i}$ has no linear term in $\xi$, the CE chain
(without the Wilson loop part) thus formed
\bea (v^{\bar i}\Theta_{\bar i},\cdots v^{\bar i}\Theta_{\bar
i})\nn\eea%
is a cycle up to $\bar\partial$-exact terms. In
ref.\cite{Rozansky:1996bq}, this is used as a weight system for
3-manifold invariant valued in $H_{\bar\partial}$. One can wedge
it with proper powers of $\Omega$ and integrate it over the
hyperK\"ahler manifold to get a complex valued 3-manifold
invariant.

If one includes the Wilson loops and take the  ``adjoint"
representation for $Q_{rw}$, a cycle in the extended CE complex is
constructed as
\bea
\sum_{n+l=N}\frac{1}{n!l}(\underbrace{v\Theta,\cdots,v\Theta}_n;\underbrace{v\Theta,\cdots,v\Theta}_l)~.\nn\eea
The path integral evaluates this cycle producing a knot invariant
valued in $H_{\bar\partial}$.

Then it was point out in ref.\cite{sawon} that from any
holomorphic vector bundle $E$ over $M$ with curvature $K$, we can
construct a representation for $Q_{rw}$. First denote the (even)
coordinate of the fibre of $E$ as $e_I$ and define the following
\bea &&(R^p_{\bar il_1\cdots l_{p+1}})^i_{\ j}=\nabla_{l_1\cdots
l_p}R_{\bar il_{p+1}\ j}^{\ \ \ \ \ i};\ \ (K^p_{\bar il_1\cdots
l_{p+1}})^I_{\ J}=\nabla_{l_1\cdots l_p}K_{\bar il_{p+1}\ J}^{\ \
\ \ \ I}~,\nn\\
&&\Theta_{\bar i}=\sum_{p=0}^{\infty}\frac{1}{(p+3)!}(R^p_{\bar
il_1\cdots l_{p+1}})^i_{\
l_{p+3}}\Omega_{il_{p+2}}\xi^{l_1}\cdots\xi^{l_{p+3}}~,\nn\\
&&(K_{\bar i})^I_{\
J}=\sum_{p=0}^{\infty}\frac{1}{(p+1)!}(K^p_{\bar il_1\cdots
l_{p+1}})^I_{\ J}\xi^{l_1}\cdots\xi^{l_{p+1}}~,\nn \eea
$\Theta$ and $K$ satisfy a neat relation
\bea&&\bar\partial_{[\bar i}\Theta_{\bar j]}=-\{\Theta_{\bar
i},\Theta_{\bar j}\}~,\nn\\
&&\bar{\nabla}_{[\bar i}K_{\bar j]}=K_{[\bar i}K_{\bar
j]}-\{\Theta_{[\bar i},K_{\bar j]}\}~.\label{neat_relation}\eea
So we can define the representation
\bea \tilde Q_{rw}=v^{\bar i}\bar{\nabla}_{\bar i}+v^{\bar
i}(\{\Theta_{\bar i},\cdot\}-(K_{\bar i})^I_{\
J}e_I\frac{\partial}{\partial e_J})~,\nn\eea
where
$$\bar\nabla_{\bar{i}} = \bar{\partial}_{\bar{i}} + (A_{\bar{i}} )^I_{~J} e_I \frac{\partial}{\partial e_J} $$
is the covariant derivative of $E$, then
$\tilde Q_{rw}^2=0$. This formula is proved in appendix.
 This agrees with our definition of the representation for
  a graded manifold equipped with homological vector field.

We can use this construction to form weight system for knots, we
write down the necessary formulae. To avoid clutter, we absorb $v$
and write $K=v^{\bar i}K_{\bar i}$ and $\Theta=v^{\bar
i}\Theta_{\bar i}$. We also raise the holomorphic indices from the
left by $\Omega^{-1}$:
$\Theta^i=(\Omega^{-1})^{ij}\Theta_j=(\Omega^{-1})^{ij}\partial_{\xi^j}\Theta$,
etc.

According to the discussion above, we form a cycle $c$,
\bea c=\frac{1}{4}(\ ;K,K,K,K)-\frac{1}{3}(
\Theta;K,K,K)+\frac{1}{4}(\Theta,\Theta;K,K)-\frac{1}{6}(\Theta,\Theta,\Theta;K)~.\nn\eea
We defined a mapping $\beta$ from the extended CE chain to graph
chain in the paragraph around Eq.\ref{beta} of the appendix.
Applying $\beta$ to the cycle $c$ we obtain (the last term in $c$
drops)
\bea\beta c&=&\frac{1}{2}\Tr(K_i,K^i,K^j,K_j)[\Gamma_1]
+\frac{1}{4}\Tr(K_i,K_j,K^i,K^j)[\Gamma_2]\nn\\
&&-\frac{1}{3}\Theta^{ijk}\Tr(K_i,K_j,K_k)[\Gamma_3]
+\frac{1}{4}\Theta^i_{\
kl}\Theta^{jkl}\Tr(K_i,K_j)[\Gamma_4]~.\label{RW_Wght}\eea
The $[\Gamma_{1,2,3,4}]$ refers to the graphs in
fig.\ref{example_Lie_wght_fig}. Of course at this low order, we
can not expect anything new from the RW weight system, because up
to degree 10, Bar-Natan has explicitly computed the dimension of
knot invariants and shown that all are covered by the Lie algebra
weight systems. Still, we need to clean up the messy expression
above. First, by using Eq.\ref{neat_relation} and dropping total
$\bar\partial$ derivatives it is straightforward to regroup the
four terms as
\bea\beta
c=-\frac{1}{4}\Tr(K_i,K^i,K_j,K^j)\big(2[\Gamma_1]+[\Gamma_2]\big)
+\frac{1}{4}\Theta^i_{\
kl}\Theta^{jkl}\Tr(K_i,K_j)\big(-\frac{2}{3}[\Gamma_3]-\frac{1}{2}[\Gamma_2]+[\Gamma_4]\big)~,\label{beta_c}\eea
and this is again a linear combination of two graph cycles as in
the case of Lie algebra weight system Eq.\ref{LieAlgWght}, except
the coefficient now takes value in $H^{\cdot}_{\bar\partial}(M)$.

\begin{figure}[h]
\begin{center}
\includegraphics[width=1in]{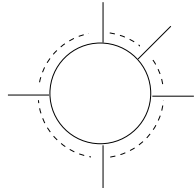}
\caption{A wheel is made of the rim and its
spokes}\label{wheel_fig}
\end{center}
\end{figure}

In general this is all we can say for the RW weight system, but
when the graph in question are made of wheels, the coefficients
can be further shuffled to be expressed in terms of characteristic
classes. The wheels are most easily defined by pictures
fig.\ref{wheel_fig},
In fig.\ref{wheel_fig}, the circle or the rim may either be the
boundary circle of the extended graph or be made of loop of
internal edges. The rim basically provides a 'trace' of the
relavent indices. It is easy to see that when the rim of
fig.\ref{wheel_fig} is the boundary circle, its weight will be
given by $\Tr[K_{i_1}K_{i_2}\cdots K_{i_n}]$, where
$K_i=\partial_{\xi^i}K|_{\xi=0}$ and $i_1,i_2,\cdots $ are the
indices of the 'spokes'. While if the rim is made of internal
edges, the weight is given by
\bea v^{\bar i_1}\cdots v^{\bar i_n}R_{\bar i_1i_1\ l_2}^{\ \ \
l_1}R_{\bar i_1i_2\ l_3}^{\ \ \ l_2}\cdots R_{\bar i_ni_n\ l_1}^{\
\ \ l_{n-1}}= v^{\bar i_1}\cdots v^{\bar i_n}\Tr[R_{\bar
i_1i_1}R_{\bar i_2i_2}\cdots R_{\bar i_ni_n}]~.\nn\eea
In both cases, if the indices $i_1,\cdots i_n$ \emph{were
anti-symmetrized}, then the weight is none other than the $n^{th}$
Chern character of the bundle $E$ and the holomorphic tangent
bundle of $M$. In general, all the Chern classes are expressible
as weights of properly chosen graphs but not vice versa (see
Sawon's discussion sec.2.2 of ref.\cite{sawon}).

With less than 4 vertices, all graphs can be formed by sewing
together the spokes of wheels. For example, $\Gamma_{1,2}$ are
made by sewing the 4 spokes of a 4-wheel while $\Gamma_4$ is made
by sewing together 2 2-wheels. So it is possible to reexpress
these weights as polynomials of Chern classes, (in fact Chern
character is more convenient here). To do this, we need a simple
but useful formula,
\bea &&(\omega^{-1})^{a_1b_1}\cdots
(\omega^{-1})^{a_pb_p}O_{a_1b_1\cdots
a_pb_p}\nn\\
&&\hspace{2cm}=e^{a_1b_1\cdots a_nb_n}\omega_{a_1b_1}\cdots
\omega_{a_{n-p}b_{n-p}}O_{a_{n-p+1}b_{n-p+1}\cdots
a_nb_n}\frac{4^p(-1)^pp!}{(n-p)!(2p)!\textrm{pf}\omega}\nn\\
&&\hspace{2cm}=e^{a_1b_1\cdots a_nb_n}(\omega^{n-p}\wedge
O)_{a_1b_1\cdots
a_nb_n}\frac{2^{n+p}(-1)^pp!}{(2n)!(n-p)!\textrm{pf}\omega}~,\label{lemma}\eea
where $\omega_{ab}$ is a symplectic form on a dim $2n$ manifold,
$O$ is a $2p$-form and $e^{a_1\cdots a_{2n}}$ is the Levi-Civita
symbol. This formula allows us to finesse away $\omega^{-1}$ form
lhs. To derive this formula, one can consider the following
fermion integral
\bea I=\int d^{2n}\psi~
e^{\psi^a\omega_{ab}\psi^b+2J_a\psi^a}\nn\eea
and evaluate it in two ways, 1. brute force, which does not
involve $\omega^{-1}$ since $\psi$ are fermions, 2. complete the
square which does involve $\omega^{-1}$. By differentiating both
sides w.r.t $J$ one obtains Eq.\ref{lemma}.

Back to our manipulation of the weights. Let us first give some
short-hands to lighten the formulae. Denote the cumbersome number
on the rhs of Eq.\ref{lemma} by $I_{\Omega}^p$. And for any form
$\tau\in\Omega^{q,q}(M)$ we use $\tau_{i_1\cdots i_q}$ to denote
${1}/{q!}v^{\bar i_1}\cdots v^{\bar i_q}\tau_{\bar i_1i_1\cdots
\bar i_qi_q}$ since the anti-holomorphic part of the forms only go
along for the ride. Finally define $\langle \tau
\rangle=e^{i_1\cdots i_{2n}}\tau_{i_1\cdots i_{2n}}$ for any
$\tau\in \Omega^{2n,2n}(M)$.

By applying Eq.\ref{lemma}, we get
\bea &&\Theta_{ikl}\Theta_j^{\
kl}\Tr[K^iK^j]=-16\pi^4ch_2(M)_{ij}ch_2(E)^{ij}\nn\\
&&\hspace{2cm}=-8\pi^4
\Big((I_{\Omega}^1)^2\langle\Omega^{n-1}ch_2(M)\rangle\langle\Omega^{n-1}ch_2(E)\rangle-I_{\Omega}^2\langle\Omega^{n-2}ch_2(M)
ch_2(E)\rangle\Big)\nn\eea
and for the 4 $K$ term
\bea
(\Omega^{-1})^{ij}(\Omega^{-1})^{kl}\Tr[K_{[i}K_jK_kK_{l]}]=4!16\pi^4(\Omega^{-1})^{ij}(\Omega^{-1})^{kl}ch_4(E)_{ijkl}
=4!16\pi^4I^2_{\Omega}\langle\Omega^{n-2}ch_4(E)\rangle~.\nn\eea
Thus Eq.\ref{beta_c} becomes
\bea\beta
c&=&-4\pi^4I^2_{\Omega}\langle\Omega^{n-2}ch_4(E)\rangle\cdot c_1~,\nn\\
&&-2\pi^4
\big((I_{\Omega}^1)^2\langle\Omega^{n-1}ch_2(M)\rangle\langle\Omega^{n-1}ch_2(E)\rangle-I_{\Omega}^2\langle\Omega^{n-2}ch_2(M)
ch_2(E)\rangle\big)\cdot(\frac{1}{6}c_1+c_2)~,\nn\\
c_1&=&2[\Gamma_1]+[\Gamma_2]~,\nn\\
c_2&=&-\frac{2}{3}[\Gamma_3]-\frac{1}{2}[\Gamma_2]+[\Gamma_4]~.\nn\eea
This is a slight strengthening of Sawon's result; in his treatment
he required the dimension of the hyperK\"ahler manifold to be
twice the number of vertices and the spokes of the wheels totally
anti-symmetrized to convert the weights into \emph{Chern numbers}.
The dimension requirement is too restrictive, for we will then not
be able to use high dimension hyperK\"ahler manifolds as weights
for low degree graphs. Here by using the cohomology group
$H_{\bar\partial}$ as coefficients, we can circumvent the
restriction.

However, in his thesis \cite{sawon}, Sawon was able to plumb much
deeper into the relation between wheels and Chern classes. By
using the so called wheeling theorem \cite{wheeling}, he managed
to identify subclasses of wheels and their weight as polynomials
of Chern classes of $TM$ and $E$: $ch(M)Td^{1/2}(E)$.

\section{Unresolved Problems}\label{UNPB}
In this last section, we discuss some loose ends and unresolved
problems.

\subsection{Vogel's Construction} This subsection serves as an
explanation why do we try to generalize the Lie algebra weight
system.

The graph complex serves as the middle man between
$\Omega^{\cdot}$(Imb) and the extended CE complex. On one side,
the mapping $c^{n,l}_{(q)}:{\cal G}_{n,l}\to \Omega^q(\Imb)$ is a
homomorphism and thus it models $H^{\cdot}(\Imb)$ on
$H^{\cdot}({\cal G})$. Even though it is not clear to us whether
this mapping is into or onto, it is shown in
ref.\cite{Cattaneo:1996pz} that this mapping does produce
infinitely many nontrivial classes in $H^{\cdot}$(Imb) for the
case $S^1\hookrightarrow \BB{R}^n$ $n>3$.

On the other side, every Lie algebra weight system produces cycles
in $H_{\cdot}({\cal G})$ through the construction
Eq.\ref{cycle_Q_R} with $\Theta$ and ${\cal R}$ given by
Eq.\ref{Lie_Q_R}. If this construction exhausted all of
$H_{\cdot}({\cal G})$, then by feeding these cycles into
$c_{(q)}^{n,l}$, we can reach the entire image of $c_{(q)}^{n,l}$.

A conjecture by Bar-Natan is that all
the weight systems come from semi-simple Lie (super)algebra with
an invariant bilinear form and finite dimensional representation.
But his conjecture was negated by Vogel, which was in fact
Bar-Natan's wish. This result calls for the need of new weight systems.
We try to review Vogel's construction, for the
paper \cite{Vogel} does not provide the most pleasant bed time
reading.

Let us specialize to the Lorentz gauge, in which the graph
cochains $c^{\cdot}_{(0)}$ only 'respond' to diagrams with
tri-valent internal vertices and uni-valent peripheral vertices
(uni-trivalent graphs). Such cochains are cocycles automatically,
since the differential $\delta$ acts on a graph by splitting a
vertex into two, each of which is at least trivalent or univalent
depending on whether the initial vertex is internal or peripheral,
but $c_{(0)}$ has only trivalent or univalent vertices to start
with. To descend to graph cohomology, we need only mod out
coboundaries, this amounts to modding out the graphs I H X (the
lhs of fig.\ref{LieAlgWght_fig}). Furthermore, the remark about
orientation in sec.\ref{Extended_Graph_Complex} says it is enough
to orient the graph by ordering the three legs at each vertex. The
flipping of the cyclic ordering at a vertex flips the sign of the
graph; this is called the AS relation. Thus all the diagrams with
tri-valent internal and uni-valent peripheral vertices, mod out by
the IHX, AS relation is nothing but a very specific representative
of the cohomology group $H^{\cdot}({\cal G})$. This is usually
denoted as ${\cal CD}_{*}$ in the literature.

\begin{figure}[h]
\begin{center}
\psfrag {=}{\small{$=$}}\psfrag {phi}{\small{$u$}}
\psfrag{a}{\scriptsize{$a$}}\psfrag{b}{\scriptsize{$b$}}
\psfrag{c}{\scriptsize{$c$}}\psfrag{d}{\scriptsize{$d$}}
\includegraphics[width=1.5in]{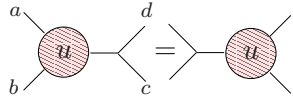}
\caption{Members of $\Lambda$ are drawn as a blob, this relation
says the insertion of this blob into any vertex does not depend on
which vertex one chooses}\label{Lambda_fig}
\end{center}
\end{figure}

Vogel defined a module structure on ${\cal CD}_{*}$, the ring for this
module is denoted $\Lambda$, which consists of trivalent graphs
with 3 external legs, satisfying the anti-symmetry under
permutation of the three legs and the relation in
fig.\ref{Lambda_fig}.
The ring multiplication is done by picking any vertex in a
uni-trivalent graph $\Gamma$, and insert a member of $\Lambda$
into it. The relation fig.\ref{Lambda_fig} says one can make
insertions into an arbitrary vertex and the result does not
change. This action is also compatible with AS and IHX relations
so it is a cohomology operation.

One can apply the Lie algebra weight system before and after the
action of $\Lambda$ and see what happens. To do this, Let $u\in
\Lambda$, $b\in H^{\cdot}({\cal G})$ and $c$ is a graph cycle
formed using the recipe of Eq.\ref{cycle_Q_R},\ref{Lie_Q_R} with
Lie algebra $g$, it turns out that
\bea (u\circ b)(c)=\chi_g(u)\times b(c)~,\nn\eea
where $u\circ b$ is the action of $u$ on $b$ and $\chi_g(u)$ is a
number called character by Vogel which only depends on the Lie
algebra and $u$. The rough reason for this simple relation is
that, the application of Lie algebra weight system to $u$ turns it
into an anti-symmetric rank 3 tensor in the Lie algebra. Letting
$e^a$ be the basis of the Lie algebra $g$, the relation
fig.\ref{Lambda_fig} is now written
\bea [e^c,u(e^a,e^b)]-u([e^a,e^b],e^c)=0~.\nn\eea
The semi-simpleness of the Lie algebra says the adjoint
representation has no non-trivial ideal, then the above relation
implies $u(e^a,e^b)$ is proportional to $[e^a,e^b]$ (Thm.6.1
\cite{Vogel}). So the effect of inserting $u$ into a trivalent
vertex is like computing vertex correction, which corrects the
tree level vertex by a factor $\chi_g(u)$--the charge
renormalization in physics.

\begin{figure}[h]
\begin{center}
\psfrag {n}{\small{$n$}}
\includegraphics[width=.8in]{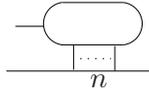}
\caption{Generating element of $\Lambda$}\label{x_n_fig}
\end{center}
\end{figure}

The hard part of the work of Vogel is to show that there is a
certain $u$ such that $\chi_g(u)=0$ for all semi-simple Lie
(super)algebra with an invariant bilinear form and finite
dimensional representation. For example, to construct $u$ such
that $\chi_{su(m)}(u)=0$, it suffices to consider $sl(m,\BB{C})$
since the latter is the complexification of the former. Apply the
$sl(m)$ weight system to $u$ will result in products like
\bea \Tr[\cdots t^d\cdots t_a\cdots t_b\cdots t_c\cdots t_d\cdots
],\nn\eea
where $t$'s are the traceless $m\times m$ matrices, the defining
representation of $sl(m,\BB{C})$. Such matrices have a very
convenient double-line notation, and hence their products can be
represented as ribbon graphs. In the case of $sl(m,\BB{C})$, the
ribbon graphs correspond to \emph{oriented} open surfaces. This
object is fact a polynomial algebra generated by the disc $t$ and
torus $\beta$. In particular, the product above will be an
oriented open surface with 3 marked points on its boundary
corresponding to $a,b,c$. It can be generated starting from a disc
with three marked points on the boundary, which is the tree level
vertex, by applying $t$ and $\beta$. This view point is important
because then for a fixed $u$, the coefficient of the polynomial of
$t,\beta$ representing $u$ \emph{does not depend on $m$}, allowing
us to define one $u$ that annihilates all $sl(m,\BB{C})$. The
effect of the disc is to create an extra index loop in the
surface, multiplying $\chi_{sl(m)}$ by $m$; while gluing the torus
does not affect $\chi_{sl(m)}$. It turns out that $x_n$ as shown
in fig.\ref{x_n_fig} for $n=2p+1$ has character (Prop 5.4, Thm
6.4, 7.1 of ref.\cite{Vogel})
\bea
\chi_{sl(m)}(x_n)=t^n+t(4\beta)^p+2^nt\beta\frac{t^{2p}-\beta^p}{t^2-\beta}\nn\eea
and $3tx_5-6t^3x_3-x_3^2+4t^6$ kills all $sl(m,\BB{C})$. The
character $\chi_{su(m)}$, however complicated, is an algebraic
function of the entries of the matrix $t^a$. The uniqueness of
analytic continuation tells us that $\chi_{su(m)}$ is the same
function of entries of $t$ as $\chi_{sl(m)}$ and hence
$\chi_{su(m)}(u)=0$. One finds a killer $u$ for each of the
semi-simple Lie algebra and multiply them together, the product
then annihilates all.

By acting $\Lambda$ on the 'Mercedes-Benz' diagram gives a map
from $\Lambda$ to $H^{\cdot}({\cal G})$. This map is shown to be
injective. In particular, we have $0\neq
u\circ(\textrm{Mercedes})$, but this cocycle will annihilate any
graph cycle constructed from Lie (super)algebras satisfying the
conditions above, since $\chi_g(u)=0$. Hence $
u\circ(\textrm{Mercedes})$ is a graph cocycle that eludes all the
Lie algebra weight systems.

We hope this sketchy review will explain why do we bother about
general weight system in this paper. But as shown by Vogel, the
member of $\Lambda$ of lowest degree that annihilates all Lie
algebra weight system has 23 vertices, so one has to come up with
constructive ways of testing the potency of Rozansky-Witten weight
system up that order, we certainly will not be able to tie up this
loose thread in this paper.

\subsection{Quantization, Surgery and Skein Relation}

In this paper we considered the properties of perturbative theory for a general
 BV-AKSZ 3D theory. However we failed to bring to light how to quantize non-perturbatively any one
of the TFT constructed in the BV-AKSZ framework. But the
quantization is crucial in the computation of the expectation
value of Wilson loop as a whole  and thus it is crucial to the
understanding of the new knot invariants arising through these
theories. As we briefly mentioned in the sec.\ref{KICST}, one pins down the
knot invariants by investigating the skein relations, which is
done in CS theory through surgeries. What one does is to take a
ball $B^3$ enclosing the locus where one strand goes over another
strand. Gouge out this ball, glue it back again after a nontrivial
diffeomorphism of the boundary of $B^3$. If the diffeomorphism is
chosen shrewdly, one finds that the previous over crossing becomes
an under crossing and the value of the knot invariant changes by a
quantity determined by the surgery. The ability of getting exact
formula for surgery in CS theory is due to the ability to quantize
this theory and to relate it to the conformal field theory. Then,
one can obtain the formula for surgery by computing the modular
transformation matrix in the corresponding CFT. For the
quantization of other AKSZ TFT's, only the case of RW theory is
known \cite{Rozansky:1996bq}. The surgery formula of RW theory is
worked out when there is no Wilson loops, so one may certainly try
to generalize the result of Rozansky and Witten to the case of
holomorphic vector bundles over hyperK\"ahler manifolds.

Without the complete knowledge of quantization, or equivalently,
the structure of the Hilbert space, one may use the path integral
to obtain (at least perturbatively) the skein relation. The
procedure will probably involve performing path integral over
$B^3$, but with arbitrary (BRST invariant) boundary conditions on
$S^2=\partial B^3$. The computation with general boundary
condition allows one to obtain surgery formula and therewith the
skein relation, regardless of what is happening outside of the
ball. This might be the point where TFT's with odd couplings, e.g.
the parameter $v^{\bar i}$ in RW theory, become interesting.

Finally let us conclude with one intriguing comment. If we
believe that our BV-AKSZ
 theories are quantum-mechanically consistent even non-perturabatively then we should
  be able to construct the corresponding knot polynomials.  In particular the theories
 with ``odd coupling constants" such
   as RW or odd Chern-Simons for integrable model (\ref{integrabmodell}) would give rise
    to knot polynomials which depend on the odd couplings and thus have
     just finite number of terms.  This idea is most intriguing for us and indeed is
      the main motivation behind our study.

%
%and as a general rule for TFT's, when we quantize a TFT on
%$\Sigma_2$, each $\Sigma_3$ that is bounded by $\Sigma_2$ gives a
%quantum state $|\psi_{\Sigma_3}\rangle$ in the Hilbert space. The
%configuration of all the fields on the boundary $\Sigma_2$ is like
%the coordinate basis $|x\rangle$, and the 'wave function'
%$\psi_{\Sigma_3}(x)=\langle x|\psi_{\Sigma_3}\rangle$ is obtained
%by doing path integral on $\Sigma_3$ with all the fields taking
%the prescribed boundary value on $\Sigma_2$. For TFT, the boundary
%configuration must be BRST invariant and hence corresponds to
%$H_Q$ as long as one can argue that only the constant modes
%contribute, which is the case of RW theory.

\section{Summary}
We summarize the major results in this paper. We have proved that
there is an isomorphism between the extended Chevalley-Eilenberg
(CE) complex of the Lie algebra of Hamiltonian vector fields with
the extended graph complex. This is an extension of the classic
result due to Kontsevich and many other authors. We also used BV
machinery to show that the path integral in a 3D TFT gives a
cocycle in the extended CE complex. Putting these two results
together, we concluded that the path integral is a cocycle of the
extended graph complex, this result generalizes that of
ref.\cite{Hamilton:2007tg}. In the BV framework, we can easily
prove that there is a homomorphism between the graph complex and
de Rham complex of the space of embeddings $S^1\to \Sigma_3$. This
so called transfer map is basically the Feynman integral.

Next, we applied these new results to bear upon the study of knot
invariants. We can form a 3D TFT associated with the Hamiltonian
lift $\Theta$ of a $Q$-structure using the AKSZ construction. The
perturbation expansion uses the Taylor coefficients of $\Theta$ as
vertex functions. This is a weight system for the extended graph
and the IHX relation is guaranteed by $\{\Theta,\Theta\}=0$, this
is a novelty compared to the Lie algebra weight system in the
sense $\{\Theta,\Theta\}=0$ takes the place of the Jaccobi
identity. The representation of the Lie algebra is replaced with
extensions of the $Q$-structure. We also showed how to construct
cycles in the extended graph complex from $Q$-structures and their
representations by applying the isomorphism of the previous
paragraph.

In these TFT's, we showed how to construct Wilson loops in
general. And the partition function or the expectation value of
the Wilson loop can be interpreted as the pairing between two dual
constructions of the extended graph complex, one from the
$Q$-structure weight system, another from the Feynman integral.
Finally, we worked out the necessary formulae for the knot
invariants with the weight system associated with a holomorphic
vector bundle on a hyperK\"ahler manifold, which is a
generalization of the Rozansky-Witten weight system.

\bigskip\bigskip

\noindent{\bf\Large Acknowledgement}:
\bigskip

\noindent It is our pleasure to thank Francesco Bonechi, Alberto
Cattaneo, Ezra Getzler and Alexei Morozov for many illuminating discussions. The
research of M.Z. was supported by VR-grant 621-2008-4273.

\bigskip\bigskip

\appendix
\section{The $Q$-Structure from Holomorphic Vector Bundle}
Let us take a GrMfld $\M=T^{0,1}[1]M$, where $M$ is K\"ahler with
coordinate $x^i,x^{\bar i}$ and $v^{\bar i}$ is the odd fibre
coordinate. There is a $Q$-structure corresponding to the
Dolbeault differential $v^{\bar i}\partial_{\bar i}$, we can find
a representation of the $Q$-structure from a holomorphic vector bundle $E\to M$.
Denote by $\nabla$ the full covariant derivative $d+\Gamma+A$ with
$\Gamma$ the Levi-Civita connection and $A$ the connection of $E$.
The holomorphicity implies we can choose the curvature $K_{\mu\nu\
J}^{\ \ I}$ of $E$ satisfy $K^{0,2}=(\bar
\partial+A^{0,1})^2=0$. Then the Biancchi identity implies for example $\nabla_{[\bar
i}K_{\bar j]k\ J}^{\ \ \;I}=0$. By applying $\nabla_{l_1}\cdots
\nabla_{\l_n}$ to the Biancchi identity, we have
\bea
0&=&\frac{1}{(n+1)!}\nabla_{l_1}\cdots\nabla_{l_n}\nabla_{[\bar
j}K_{\bar i]l_{n+1}\ K}^{\ \ \ \ \;J}+\textrm{perm in }l=\nn\\%
&=&\sum_{k=0}^{n-1}\frac{1}{(n+1)!}\nabla_{l_1\cdots
l_k}\Big[K_{l_{k+1}\bar j\ M}^{\ \ \ \ \ J}\nabla_{l_{k+2}\cdots
l_n}K_{\bar il_{n+1}\ K}^{\ \ \ \ M}-R_{l_{k+1}\bar j\ l_{k+2}}^{\
\ \ \ \
m}\nabla_{ml_{k+3}\cdots l_n}K_{\bar il_{n+1}\ K}^{\ \ \ \ J}(n-k)\nn\\%
&&-K_{l_{k+1}\bar j\ K}^{\ \ \ \ \ M}\nabla_{l_{k+2}\cdots
l_n}K_{\bar il_{n+1}\ M}^{\ \ \ \
J}\Big]\nn\\
&&+\frac{1}{(n+1)!}\nabla_{\bar
j}\nabla_{l_1}\cdots\nabla_{l_n}K_{\bar il_{n+1}\ K}^{\ \ \ \
\;J}+\textrm{perm in $l$, anti in }[\bar i,\bar j]~.\nn\eea
Define some short hands
\bea &&(R^p_{\bar il_1\cdots l_{p+1}})^i_{\ j}=\nabla_{l_1\cdots
l_p}R_{\bar il_{p+1}\ j}^{\ \ \ \ \ i}~;\ \ (K^p_{\bar il_1\cdots
l_{p+1}})^I_{\ J}=\nabla_{l_1\cdots l_p}K_{\bar il_{p+1}\ J}^{\ \
\ \ \ I}~,\nn\eea
note that all $l$'s are symmetric due to the K\"ahler property.
Continue on
{\small\bea&&\sum_{k=0}^{n-1}\sum_{p=0}^k\frac{1}{(n+1)!}C_k^p\Big[-\big[K^p_{\bar
jl_1\cdots l_{p+1}},K^{n-p-1}_{\bar il_{p+2}\cdots
l_{n+1}}\big]^J_{\ K}+(R^p_{\bar jl_1\cdots l_{p+1}})^m_{\
l_{p+2}}(K^{n-p-1}_{\bar iml_{p+3}\cdots l_{n+1}})^J_{\
K}(n-k)\Big]\nn\\
&&+\frac{1}{(n+1)!}\nabla_{\bar j}(K^n_{\bar il_1\cdots
l_{n+1}})_{\ K}^J+\textrm{perm in $l$, anti in }[\bar i,\bar j]\nn\\
&=&\sum_{p=0}^{p=n-1}\bigg[-\frac{1}{(n+1)!}\frac{n!}{(n-p-1)!(p+1)!}\big[K^p_{\bar
jl_1\cdots l_{p+1}},K^{n-p-1}_{\bar il_{p+2}\cdots
l_{n+1}}\big]^J_{\ K}\nn\\
&&+\frac{1}{(p+2)!(n-p-1)!}(R^p_{\bar jl_1\cdots l_{p+1}})^m_{\
l_{p+2}}(K^{n-p-1}_{\bar iml_{p+3}\cdots l_{n+1}})^J_{\ K}\bigg]\nn\\
&&+\frac{1}{(n+1)!}\nabla_{\bar j}(K^n_{\bar il_1\cdots
l_{n+1}})_{\ K}^J+\textrm{perm in $l$, anti in }[\bar i,\bar j]\nn\\
&=&\sum_{p=0}^{p=n-1}-\frac{1}{2(n-p)!(p+1)!}\big[K^p_{\bar
jl_1\cdots l_{p+1}},K^{n-p-1}_{\bar il_{p+2}\cdots l_{n+1}}\big]^J_{\ K}\nn\\%
&&+\frac{1}{(p+2)!(n-p-1)!}(R^p_{\bar jl_1\cdots l_{p+1}})^m_{\
l_{p+2}}(K^{n-p-1}_{\bar iml_{p+3}\cdots l_{n+1}})^J_{\
K}\nn\\
&&+\frac{1}{(n+1)!}\nabla_{\bar j}(K^n_{\bar il_1\cdots
l_{n+1}})_{\ K}^J+\textrm{perm in $l$, anti in }[\bar i,\bar
j]~.\label{lastline}\eea}
So if we define
\bea &&R_{\bar i}=\sum_{p=0}^{\infty}\frac{1}{(p+2)!}(R^p_{\bar
il_1\cdots l_{p+1}})^i_{\
j}\xi^{l_1}\cdots\xi^{l_{p+1}}\xi^j\partial_{\xi^i}~,\nn\\%
&&(K_{\bar i})^I_{\
J}=\sum_{p=0}^{\infty}\frac{1}{(p+1)!}(K^p_{\bar il_1\cdots
l_{p+1}})^I_{\ J}\xi^{l_1}\cdots\xi^{l_{p+1}}~,\nn \eea
Then the above relation can be written concisely%
\bea \nabla_{[\bar i}K_{\bar j]}=K_{[\bar i}K_{\bar j]}-R_{[\bar
i}K_{\bar j]}~.\nn\eea
So one can  place $K$'s on the Wilson loop by tracing the $I,J$
index. And the above relation says that the diagrams I+H+X (see
fig.\ref{LieAlgWght_fig}, the lower edge of figure I and the vertices of H X are now on the Wilson loop) is something $\bar\partial$-exact.

If furthermore, $M$ is also hyperK\"ahler, with symplectic form
$\Omega_{ij}$ we can define
\bea &&\Theta_{\bar
i}=\sum_{p=0}^{\infty}\frac{1}{(p+3)!}(R^p_{\bar il_1\cdots
l_{p+1}})^i_{\
l_{p+3}}\Omega_{il_{p+2}}\xi^{l_1}\cdots\xi^{l_{p+3}}~,\nn\eea
We obtain
\bea \nabla_{[\bar i}K_{\bar j]}=K_{[\bar i}K_{\bar
j]}-\{\Theta_{[\bar i},K_{\bar j]}\}~.\nn\eea
If $E$ is the tangent bundle of $M$, then we recover our previous
result
\bea\nabla_{[\bar i}\Theta_{\bar j]}=\partial_{[\bar
i}\Theta_{\bar j]}=-\frac{1}{2}\{\Theta_{[\bar i},\Theta_{\bar
j]}\}~.\nn\eea

\section{Isomorphism between Extended Graph and CE Complex}
We prove that this graph complex is isomorphic to the following
generalized CE complex. The complex is spanned by%
\bea c_{n,l}&=&\BB{X}_{f_0}\wedge\cdots\wedge\BB{X}_{f_n}\otimes
(g_0\otimes\cdots\otimes g_l)\Rightarrow (f_0,\cdots,f_n;g_0,\cdots g_l)~.\nn\eea%
We allow cyclic permutations in the second factor, with
$l+1\equiv0$, etc. The differential is defined as in
Eq.\ref{CE_diff}. The differential commutes with the $osp$ action,
so it makes sense to consider the $osp$ co-invariants of the CE
complex, that is the orbits of the $osp$ action.

Before proceeding to the proof, we temporarily change the
labelling of chains from $0\cdots n,\;0\cdots l$ to $1\cdots
n,\;1\cdots l$ and offer our apology for this confusion.

We first re-prove the known result Eq.\ref{iso_Lie_graph} to set
the stage\footnote{A version of the proof in the $\BB{R}^{2n}$
case may be found in ref.\cite{ConantVogtmann}, and
ref.\cite{Hamilton-2005} contains the $\BB{R}^{2n|m}$ case but is
rather sketchy in showing the homomorphism}. Once this is clear,
the generalization is straightforward. Most of the labor comes
from keeping track of signs, so we use the following technical
contraptions. Let $x^p$ denote the coordinate of $\BB{R}^{2n|m}$,
with the (trivial) symplectic structure
$\Omega=1/2\Omega_{pq}dx^p\wedge dx^q$. Enlarge the set of
coordinates into the $\{x_i^p,\;i=0,1\cdots\}$, $x^p_i$ for
different $i$ are the same as $x^p$, but the label $i$ allows us
to treat them as formally independent variables. Introduce some
formal degree 1 variables $\{t_i,\; i=0,1\cdots\}$. This way, we
are able to define a Laplacian which induces the Poisson bracket,
even though strictly speaking, even brackets are not induced by
Laplacians. We can write
\bea (f_1,\cdots
f_n)=\BB{X}_{f_1}\wedge\cdots\wedge\BB{X}_{f_n}\Rightarrow\
t_1f_1(x_1)\cdot t_2f_2(x_2)\cdots t_nf_n(x_n)~,\label{techno}\eea
where the odd parameter $t$ takes care of the grading shift due to
$\deg\BB{X}_f=\deg f+1$.

The Laplacian is defined somewhat awkwardly%
\bea \Delta:=-\sum_{i<j}R_{i,j}\frac{\partial}{\partial
t_j}(\Omega^{-1})^{pq}\frac{\partial}{\partial
x_i^p}\frac{\partial}{\partial x^q_j}~,\nn\eea%
where $R_{i,j}$ renames $x_j,t_j$ as $x_i,t_i$. It is easy to
check the following
{\small\bea \Delta t_mf(x_m)t_ng(x_n)&=&-(-1)^{x^qf+x^p}
R_{m,n}\frac{\partial}{\partial
t_n}(\Omega^{-1})^{pq}(t_m\partial_pf(x_m))(t_n\partial_qg(x_n))\nn\\%
&=&-(-1)^{x^qf+x^p+1+x^p+f+fx^p} t_m\{f(x_m),g(x_m)\}=(-1)^f
t_m\{f(x_m),g(x_m)\}~.\nn\eea}
This relation mimics the second one of Eq.\ref{useful}. We need to
check $\Delta^2=0$ to ensure $\Delta$ induces a differential,
\bea\Delta^2=\sum_{i<j;k<l}R_{k,l}\frac{\partial}{\partial
t_l}(\Omega^{-1})^{rs}\frac{\partial}{\partial
x^r_k}\frac{\partial}{\partial
x^s_l}R_{i,j}\frac{\partial}{\partial
t_j}(\Omega^{-1})^{pq}\frac{\partial}{\partial
x^p_i}\frac{\partial}{\partial x^q_j}~.\nn\eea%
When $k,l,i,j$ are not the same, this certainly vanishes, due to
the antisymmetry of $t$. However we may also have the following
possibilities%
\bea l=i~;\ \ \Delta^2\Rightarrow R_{k,ij}\frac{\partial}{\partial
t_i}(\Omega^{-1})^{rs}\frac{\partial}{\partial
x^r_k}\big[\frac{\partial}{\partial
x^s_i}+\frac{\partial}{\partial
x^s_j}\big]\frac{\partial}{\partial
t_j}(\Omega^{-1})^{pq}\frac{\partial}{\partial
x^p_i}\frac{\partial}{\partial x^q_j}~,\nn\\%
k=i~;\ \ \Delta^2\Rightarrow R_{i,jl}\frac{\partial}{\partial
t_l}(\Omega^{-1})^{rs}\big[\frac{\partial}{\partial
x^r_i}+\frac{\partial}{\partial
x^r_j}\big]\frac{\partial}{\partial x^s_l}\frac{\partial}{\partial
t_j}(\Omega^{-1})^{pq}\frac{\partial}{\partial
x^p_i}\frac{\partial}{\partial x^q_j}~.\nn\eea%
The third term is zero by itself. The fourth one depending whether
$j<l$ or $j>l$ can be reshuffled and cancel the first and second
term respectively.%

The homomorphism from a CE chain to the graph chain is defined as:
fix $n$ vertices labelled $1\cdots n$ to correspond to $n$ dummy
variables $x_i$; every way of connecting $n$ vertices with edges
gives a graph $\Gamma_{1,\cdots n}$. Here the subscript denotes
the dummy variable $x_1\cdots x_n$. Since the vertices of the
graph are already ordered, we need only choose an orientation of
all the edges to fix the orientation of graph. For an edge that
runs from vertex $i$ to $j$, form the operator
$(\Omega^{-1})^{pq}\partial_{x^p_i}\partial_{x^q_j}$. Doing the
same for every edge in $\Gamma_{1\cdots k}$ gives the operator
$\beta_{\Gamma_{1,\cdots k}}$. A graph like the following will
\begin{figure}[h]
\begin{center}
\psfrag {v1}{\small{$1$}}\psfrag {v2}{\small{$2$}}\psfrag {v3}{\small{$3$}}\psfrag{v4}{\small{$4$}}%
%\psfrag {i1}{$i_1$}\psfrag {i2}{$i_2$}\psfrag{i3}{$i_3$}%\psfrag{i4}{$i_4$}%
%\psfrag {j1}{$j_1$}\psfrag{j2}{$j_2$}\psfrag{j3}{$j_3$}%\psfrag{j4}{$j_4$}%
%\psfrag{k1}{$k_1$}\psfrag {k2}{$k_2$}\psfrag {k3}{$k_3$}%\psfrag{k4}{$k_4$}%
%\psfrag {l1}{$l_1$}\psfrag {l2}{$l_2$}\psfrag{l3}{$l_3$}%\psfrag{l4}{$l_4$}%
\psfrag {i1}{}\psfrag {i2}{}\psfrag {i3}{}\psfrag {j1}{}\psfrag
{j2}{}\psfrag {j3}{}\psfrag {k1}{}\psfrag {k2}{}\psfrag
{k3}{}\psfrag {l1}{}\psfrag {l2}{} \psfrag {l3}{}
\includegraphics[width=.8in]{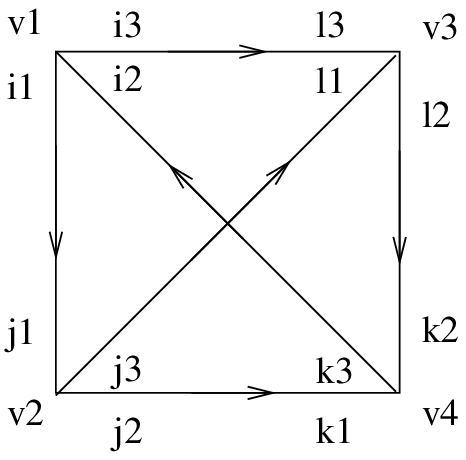}
\caption{Box}\label{Box_fig}
\end{center}
\end{figure}
form the operator
{\small\bea &&\beta_{\Gamma}=
((\Omega^{-1})^{p_3q_3}\frac{\partial}{\partial
x^{p_3}_4}\frac{\partial}{\partial x^{q_3}_1})%
((\Omega^{-1})^{p_1q_1}\frac{\partial}{\partial
x^{p_1}_1}\frac{\partial}{\partial x^{q_1}_2})%
((\Omega^{-1})^{p_2q_2}\frac{\partial}{\partial
x^{p_2}_3}\frac{\partial}{\partial x^{q_2}_4})\nn\\%
&&\hspace{2.3cm}((\Omega^{-1})^{r_2s_1} \frac{\partial}{\partial
x^{r_2}_2}\frac{\partial}{\partial x^{s_1}_3})%
((\Omega^{-1})^{r_3s_2}\frac{\partial}{\partial
x^{r_3}_4}\frac{\partial}{\partial x^{s_2}_1})%
((\Omega^{-1})^{r_3s_1}\frac{\partial}{\partial
x^{r_3}_2}\frac{\partial}{\partial x^{s_1}_4}~.\nn\eea}

One immediately realizes that an edge that forms a loop will be
associated with the operator
$(\Omega^{-1})^{pq}\partial_{x_i^p}\partial_{x_i^q}=0$, which is
in accordance with the observation in
sec.\ref{Extended_Graph_Complex} that graphs with loops are zero.
Next we define homomorphism taking a CE chain $c$ to the following
\bea \beta c=\sum_{\Gamma_{1\cdots k}}[\Gamma_{1\cdots k}]\int
dt_{k}\cdots dt_{1}\cdot\beta_{\Gamma_{1\cdots
k}}c\Big|_{x=0}~,\label{beta}\eea
where the sum is over \emph{all possible ways of connecting $n$
vertices}. The way this operator acts on the CE chain formally
resembles the Wick formula for Gaussian integral. If the reader
had been wondering about the seeming loss of the graded
commutativity in Eq.\ref{techno}, the sum of all ways of
connecting is the remedy. As an example, take
$c=t_1f_2(x_1)t_2f_1(x_2)\cdots=(-1)^{(f_1+1)(f_2+1)}t_2f_1(x_2)t_1f_2(x_1)\cdots$.
Since the $t_{1,2},x_{1,2}$ are dummy variables, we can rename
them in $c$ and in $\beta$ simultaneously. The resulting graph
will be
$(-1)^{(f_1+1)(f_2+1)}\beta(t_1f_1(x_1)t_2f_2(x_2)\cdots)$, as it
should be.

Next we show that this is a homomorphism: $\beta
\partial c=\partial \beta c$. We investigate $\partial_I$ first,
since this one also appears independently in the un-extended
CE/graph complex. Take $c=t_1f_1(x_1)\cdots
t_{n+1}f_{n+1}(x_{n+1})$,
\bea
\partial_I c&=&\sum_{1\leq i<j\leq n+1}c_{ij}\nn\\
c_{ij}&=&(-1)^{s_{ij}} t_0\{f_i,f_j\}(x_0)t_1f_1(x_1)\cdots \hat
i\cdots \hat j\cdots t_{n+1}f_{n+1}(x_{n+1})\nn\\
\beta c_{ij}&=&\sum_{\Gamma}[\Gamma_{0,\cdots\hat i,\cdots \hat
j,\cdots n+1}]\int dt_{n+1}\cdots \hat{dt_j}\cdots
\hat{dt_i}\cdots dt_0\ \beta_{\Gamma_{0,\cdots\hat i,\cdots \hat
j,\cdots n+1}}\ c_{ij}\big|_{x=0}~,\nn\eea
where $\Gamma_{0,\cdots\hat i,\cdots \hat j,\cdots n+1}$ means the
$n$ vertices in this graph correspond to the dummy variable
$x_0,\cdots \hat{x}_i,\cdots \hat{x}_j,\cdots x_n$.

\begin{figure}[h]
\begin{center}
\psfrag {0}{\scriptsize{$0$}}\psfrag
{i}{\scriptsize{$i$}}\psfrag{j}{\scriptsize{$j$}}\psfrag
{+}{\scriptsize{$+$}}
\includegraphics[width=2.7in]{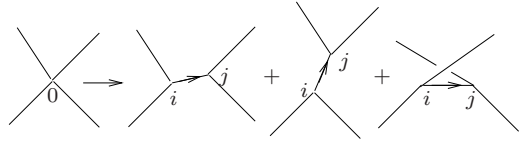}
\caption{Breaking into two internal vertices}\label{Breaking_fig}
\end{center}
\end{figure}

Variable $x_0$ will appear among the differentials of
$\beta_{\Gamma_{0,\cdots\hat i,\cdots \hat j,\cdots n+1}}$, but if
we agree to understand the following
\bea (\Omega^{-1})^{pq}\frac{\partial}{\partial
x_0^p}\frac{\partial}{\partial
x_k^q}\rightarrow(\Omega^{-1})^{pq}\big(\frac{\partial}{\partial
x_i^p}+\frac{\partial}{\partial
x_j^p}\big)\frac{\partial}{\partial x_k^q}~,\nn\eea%
for every $x_0$ in $\beta_\Gamma$, we can rewrite $\beta c_{ij}$
as%
\bea\beta c_{ij} &=&\sum_{\Gamma}[\Gamma_{0,\cdots\hat i,\cdots
\hat j,\cdots n+1}]\int dt_{n+1}\cdots \hat{dt_j}\cdots
\hat{dt_i}\cdots dt_1dt_0\ \beta_{\Gamma_{0,\cdots\hat i,\cdots
\hat j,\cdots n+1}}\nn\\%
&&R_{0,ij}(-1)\frac{\partial}{\partial
t_j}(\Omega^{-1})^{pq}\frac{\partial}{\partial
x^p_i}\frac{\partial}{\partial x^q_j}
\big(t_1f_1(x_1)\cdots t_{n+1}f_{n+1}(x_{n+1})\big)\nn\\%
&=&\sum_{\Gamma,\hat{\Gamma}}[\Gamma_{0,\cdots\hat i,\cdots \hat
j,\cdots n+1}]\int dt_{n+1}\cdots dt_1\
\beta_{\hat{\Gamma}_{1,\cdots
n+1}}(-1)^{i+j-1}\big(t_1f_1(x_1)\cdots
t_{n+1}f_{n+1}(x_{n+1})\big)\big|_{x=0}~,\nn\eea
where $\hat{\Gamma}$ is obtained from $\Gamma$ by breaking up the
vertex 0 in $\Gamma$ into vertex $i,j$ ($i<j$) in all possible
ways as in fig.\ref{Breaking_fig}. Since we assumed that all $f$
are at least cubic, the uni- or bi-valent vertex will not occur in
the breaking.
The sign factor $(-1)^{i+j+1}$ is taken to be the
\emph{definition} of the incidence number between
$[\Gamma_{0,\cdots\hat i,\cdots \hat j,\cdots n+1}]$ and
$[\hat{\Gamma}_{1,\cdots n+1}]$. This means the last line is
exactly $\partial_I\beta c$ and we have re-produced the proof of
ref.\cite{ConantVogtmann,Hamilton-2005}. The discrepancy of the
incidence number here with that of fig.\ref{sign_fac_fig} is due
to two reasons, there we combined vertices $i,j$ and named it $i$
instead of 0 and also the labelling of the vertices there stars
from 0 instead of 1. These two factors causes
$(-1)^{i+j+1}\to(-1)^{i+j+1+(i-1)}\to(-1)^{i+(j+1)+1+(i-1)}=(-1)^{j+1}$.

To proceed to the extended graph, we use extra formal odd
parameters $t_{n+1}\cdots t_{n+l}$; this way we can write the
entire chain as one function
\bea (f_1,\cdots,f_n;g_1,\cdots g_l)=t_1f_1(x_1)\cdots
t_nf_n(x_n)t_{n+1}g_1(x_{n+1})\cdots
t_{n+l}g_l(x_{n+l})~.\nn\eea%
One need not worry about seeming loss of cyclicity here either.

\begin{figure}[h]
\begin{center}
\psfrag {j}{\scriptsize{$j$}}\psfrag {i}{\scriptsize{$i$}}\psfrag
{+}{\scriptsize{$+$}}
\includegraphics[width=2.5in]{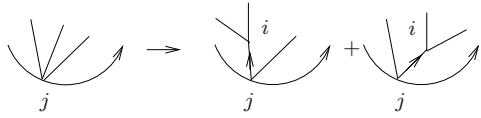}
\caption{Breaking into an internal and a peripheral
vertex}\label{Breaking2_fig}
\end{center}
\end{figure}

To construct the operator, let $\Gamma_{1,\cdots n;1\cdots l}$
denote an extended graph. We form $\beta$ like before, but also
for every edge that runs from a peripheral vertex to an internal
one, we include an operator
\bea (\Omega^{-1})^{pq}\frac{\partial}{\partial
x^p_{n+j}}\frac{\partial}{\partial x^q_{i}}~;\ \ 1\leq j\leq l,
1\leq i \leq n~,\nn\eea
and for every edge that runs from vertex $i$ to $j$ on the Wilson
loop, we include
\bea (\Omega^{-1})^{pq}\frac{\partial}{\partial
x^p_{n+i}}\frac{\partial}{\partial x^q_{n+j}}~;\ \ 1\leq i,j\leq l~.
\nn\eea
To investigate $\partial_V$, let $c=(f_1,\cdots f_n;g_1,\cdots
g_l)$,
\bea \partial_Vc&=&\sum_{1\leq i\leq n;1\leq j\leq l}c_{ij}~,\nn\\%
c_{ij}&=&-(-1)^{t_{ij}}(t_1f_1(x_1))\cdots\widehat{(t_if_i(x_i))}\cdots
(t_nf_n(x_n))\nn\\%
&&\hspace{2cm}\big( t_{n+1}g_1(x_{n+1})\cdots
t_{n+j}\{f_i,g_j\}(x_{n+j})\cdots t_{n+l}g_l(x_{n+l})~,\nn\\%
\beta c_{ij}&=&\sum_{\Gamma}[\Gamma_{1,\cdots\hat i,\cdots
n;1\cdots l}]\int dt_{n+l}\cdots dt_{n+1}dt_n\cdots
\hat{dt_i}\cdots dt_1\ \beta_{\Gamma_{1,\cdots\hat i,\cdots
n;1\cdots l}}c_{ij}\big|_{x=0}~.\nn\eea
The only tricky terms in $\beta_{\Gamma}$ are of the type%
\bea (\Omega^{-1})^{pq}\frac{\partial}{\partial
x^p_{n+j}}\frac{\partial}{\partial x^q_{k}},\ k\neq i~.\nn\eea%
But if we agree to replace \bea &&\frac{\partial}{\partial
x_{n+j}} \Rightarrow\ \ \frac{\partial}{\partial
x_{n+j}}+\frac{\partial}{\partial x_j}\nn\eea%
within $\beta_{\Gamma}$, then $\beta c_{ij}$ can be written as
\bea \beta c_{ij}&=&\sum_{\Gamma}[\Gamma_{1,\cdots\hat i,\cdots
n;1\cdots
l}]\int dt_{n+l}\cdots \hat{dt_i}\cdots dt_1\nn\\%
&& R_{n+j,i}\beta_{\Gamma_{1,\cdots\hat i,\cdots n;1\cdots
l}}\Big(- \frac{\partial}{\partial
t_i}(\Omega^{-1})^{rs}\frac{\partial}{\partial
x^r_{n+j}}\frac{\partial}{\partial x^s_{i}}\Big)(f_1,\cdots
f_n;g_1,\cdots g_l)\nn\\%
&=&\sum_{\Gamma,\hat{\Gamma}}[\Gamma_{1,\cdots\hat i,\cdots
n;1\cdots l}]\int dt_{n+l}\cdots dt_1 (-1)^i
\beta_{\hat{\Gamma}_{1,\cdots n;1\cdots
l}}(f_1,\cdots f_n;g_1,\cdots g_l)\big|_{x=0}~,\nn\eea
where $\hat\Gamma$ is obtained from $\Gamma$ by splitting the
$j$'th peripheral vertex into an internal one and an peripheral
one as in fig.\ref{Breaking2_fig}.
And $(-1)^i$ is defined to be the incident number between
$[\Gamma_{1,\cdots\hat i,\cdots n;1\cdots l}]$ and
$[\hat{\Gamma}_{1,\cdots n;1\cdots l}]$. So we have again
$\beta\partial_V c=\partial_V\beta c$.

\begin{figure}[h]
\begin{center}
\psfrag {j}{\scriptsize{$j$}}\psfrag {k}{\scriptsize{$j+1$}}
\includegraphics[width=2.3in]{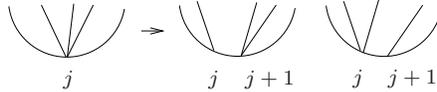}
\caption{Breaking into two adjacent peripheral
vertices}\label{Breaking3_fig}
\end{center}
\end{figure}

Now look at $\partial_H$,
\bea
\partial_H c&=&\sum_{1\leq j< l}c_j+c_l~,\nn\\%
c_j&=&-(-1)^{\sum_1^n(f_k+1)+\sum_1^j(g_k+1)}(t_1f_1(x_1))\cdots
(t_nf_n(x_n))\nn\\
&&t_{n+1}g_1(x_{n+1})\cdots t_{n+j}g_jg_{j+1}(x_{n+j})
t_{n+j+2}g_{j+2}(x_{n+j+2})\cdots
t_{n+l}g_l(x_{n+l})~,\ \ 1\leq j<l~,\nn\\
c_l&=&-(-1)^{\sum_1^n(f_k+1)+(g_l+1)\sum_1^l(g_k+1)}(t_1f_1(x_1))\cdots
(t_nf_n(x_n))\nn\\
&&t_{n+1}g_lg_1(x_{n+1})\cdots t_{n+l-1}g_{l-1}(x_{n+l-1})~,\nn\\
\beta c_j&=&\sum_{\Gamma}[\Gamma_{1,\cdots
n;1\cdots\widehat{j+1},\cdots l}]\int dt_{n+l}\cdots
\hat{dt}_{n+j+1}\cdots dt_1\ R_{n+j,n+j+1}\beta_{\Gamma_{1,\cdots
n;1\cdots\widehat{j+1},\cdots l}}\ c_j\big|_{x=0}~,\nn\eea
For this term, we only need to worry about operators like
\bea (\Omega^{-1})^{pq}\frac{\partial}{\partial
x^p_{n+j}}\frac{\partial}{\partial x^q_{k}}~,\textrm{or }~
(\Omega^{-1})^{pq}\frac{\partial}{\partial
x^p_{n+j}}\frac{\partial}{\partial x^q_{n+i}},~~i\neq j+1\nn\eea
But if we agree to replace%
\bea &&\frac{\partial}{\partial x_{n+j}} \Rightarrow\ \
\frac{\partial}{\partial x_{n+j}}+\frac{\partial}{\partial
x_{n+j+1}}~,\nn\eea
within $\beta_{\Gamma}$, then we can write $\beta c_j,~j<l$ as
\bea \beta c_j&=&-\sum_{\Gamma}[\Gamma_{1,\cdots
n;1\cdots\widehat{j+1},\cdots l}]\int dt_{n+l}\cdots
\hat{dt}_{n+j+1}\cdots dt_1\nn\\
&&\hspace{1.5cm}R_{n+j,n+j+1}\beta_{\Gamma_{1,\cdots n;1\cdots l}
} \frac{\partial}{\partial
t_{n+j+1}}(f_1,\cdots f_n;g_1,\cdots g_l)\big|_{x=0}\nn\\%
&=&-\sum_{\Gamma,\hat{\Gamma}}[\Gamma_{1,\cdots
n;1\cdots\widehat{j+1},\cdots l}]\int dt_{n+l}\cdots dt_1\
(-1)^{n+j}\beta_{\hat{\Gamma}_{1,\cdots n;1\cdots l} }(f_1,\cdots
f_n;g_1,\cdots g_l)\big|_{x=0}~,\nn\eea
where $\hat{\Gamma}$ is obtained by breaking vertex $j$ in
$\Gamma$ into vertex $j$ and $j+1$ in all possible ways as in
fig.\ref{Breaking3_fig}.
Naturally, we take $(-1)^{n+j+1}$ as the incidence number between
$[\Gamma_{1,\cdots n;1\cdots\widehat{j+1},\cdots l}]$ and
$[\hat{\Gamma}_{1,\cdots n;1\cdots l}]$. One special case which we
do not cover is when $j=l$, but the reader can easily verify that
$\partial_H$ combines vertex $l$ and $1$ and names the new vertex
1 with incident number $(-1)^{n+l-1}$. The final conclusion is
that $\beta\partial_Hc=\partial_H\beta c$ and we completed the
proof that $\beta$ induces a chain map between the extended CE
complex and the extended graph complex.

\begin{figure}[h]
\begin{center}
\psfrag {v1}{}\psfrag {v2}{}\psfrag {u3}{}\psfrag {u1}{}\psfrag
{u2}{}\psfrag {u4}{}
%
%\psfrag {v1}{\scriptsize{$\underline{1}$}}\psfrag
%{v2}{\scriptsize{$\underline{2}$}} \psfrag
%{u3}{\scriptsize{$\underline{3}$}}\psfrag
%{u1}{\scriptsize{$\underline{1}$}}\psfrag{u2}{\scriptsize{$\underline{2}$}}\psfrag
%{u4}{\scriptsize{$\underline{4}$}}
%
\psfrag
{e1}{\scriptsize{$1$}}\psfrag{e2}{\scriptsize{$2$}}\psfrag{e3}{\scriptsize{$3$}}
\psfrag{e4}{\scriptsize{$4$}}\psfrag{e5}{\scriptsize{$5$}}
\includegraphics[width=0.8in]{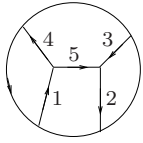}
\caption{Example of the inverse map, the vertices are labelled as:
1 and 2 for the left and right internal vertex, 1 for the lower
left peripheral vertex and numbering increases
counterclockwise}\label{dec_graph_inver_fig}
\end{center}
\end{figure}

Next we construct the inverse of $\beta$. Take a super vector
space $\BB{R}^{2n|m}$, $n,m$ big enough (bigger than the number of
edges in a graph). Let $p_i,q_i,\ i=1\sim n$ be the bosonic
coordinates with the bracket
$\{p_i,q_j\}=-\{q_j,p_i\}=\delta_{ij}$. Let us number the edges
from 1 through $k$, for an edge $i$, we associate $p_i$ to the
vertex from which it issues forth and $q_i$ to the vertex on which
it ends. This way, we form a polynomial for each
vertex. For example, the fig.\ref{dec_graph_inver_fig} gives
\bea c=(q_1p_5p_4,q_5p_2q_3;p_1, q_2, p_3, q_4)~.\label{inverse}\eea

Clearly $\beta$ acting on $c$ only produces one nonzero term which
is exactly the graph we want. It is perhaps queer that only the
bosonic part of $\BB{R}^{2n|m}$ is used in this construction, but
one must remember that $c$ is to be regarded as a representative
of the orbit under the $osp_{2n|m}$ action. One can also use the
odd part of $\BB{R}^{2n|m}$ to form the same inverse. But this
time, it is more convenient to use the alternative orientation
scheme for the graph complex, one orders all the even valent
vertices and orders all incident legs of every vertex\footnote{See
the remark in the second paragraph of
sec.\ref{Extended_Graph_Complex}}. Let $\xi^i$ be the odd
coordinates, even though we can no longer choose $p$ or $q$ make
distinction between income or outgoing legs, we can place the
$\xi$'s in one vertex in the order conforming to the ordering of
the incident legs for that vertex.

\begin{figure}[h]
\begin{center}
\psfrag {0}{\scriptsize{$0$}}\psfrag
{1}{\scriptsize{$1$}}\psfrag{2}{\scriptsize{$2$}}\psfrag
{3}{\scriptsize{$3$}}\psfrag {A}{\small{$\Gamma_1$}}\psfrag
{B}{\small{$\Gamma_2$}}\psfrag {C}{\small{$\Gamma_3$}}\psfrag
{D}{\small{$\Gamma_4$}}
\includegraphics[width=2.8in]{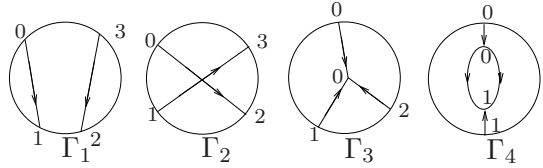}
\caption{Lowest order cycles}\label{example_Lie_wght_fig}
\end{center}
\end{figure}

Finally as an exercise, let us see an example of graph cycles
constructed from the CE cycle. Let $A^{\alpha}$ be the odd
coordinate of $su(n)[1],so(n)[1],sp(2n)[1]$,
$\eta_{\alpha\beta}=\Tr[T_{\alpha}T_{\beta}]$ be the killing
metric and also let
\bea c&=&\frac{1}{4}(\
;A,A,A,A)+\frac{1}{3}(\frac{1}{3}\Tr[A^3];A,A,A)\nn\\
&&+\frac{1}{2!2}(\frac{1}{3}\Tr[A^3],
\frac{1}{3}\Tr[A^3];A,A)+\frac{1}{3!}(\frac{1}{3}\Tr[A^3],
\frac{1}{3}\Tr[A^3],\frac{1}{3}\Tr[A^3];A)~,\nn\eea
where $A=A^{\alpha}T_{\alpha}$. One can check directly that $c$ is
closed. Applying $\beta$, we get the four graphs in
fig.\ref{example_Lie_wght_fig},
with coefficients (where $T^{\alpha}=\eta^{\alpha\beta}T_{\beta}$)
\bea \beta
c&=&-\frac{1}{2}\Tr[T^{\alpha}T_{\alpha}T_{\beta}T^{\beta}][\Gamma_1]-\frac{1}{4}
\Tr[T^{\alpha}T^{\beta}T_{\alpha}T_{\beta}][\Gamma_2]-\frac{i}{3}f_{\alpha\beta\gamma}
\Tr[T^{\alpha}T^{\beta}T^{\gamma}][\Gamma_3]-\frac{1}{4}f_{\alpha\beta\gamma}f^{\alpha\beta\gamma}[\Gamma_4]\nn\\
&=&\frac{d_GC_2(G)}{2}\big(\frac{1}{4}[\Gamma_2]+\frac{1}{3}[\Gamma_3]-\frac{1}{2}[\Gamma_4]\big)
-\frac{d_rC_2^2(r)}{4}\big([\Gamma_2]+2[\Gamma_1]\big)~,\label{LieAlgWght}\eea
where $d_{r,G}$ is the dimension of representation $r$ and adjoint
representation, $C_2$ is the second Casimir. So the result is the
linear combination of two cycles in the two brackets above. Just
for a record, the coefficient for the first cycle is
$n^2(n^2-1),n(n-1)(n-2),n(2n+1)(n+1)$ and
$-(n^2-1)^2/(4n),-n(n-1)^2,-n(2n+1)/8$ for the second.


\begin{thebibliography}{1}

\newcommand{\np}{{\em Nucl.\ Phys.\ }}
\newcommand{\pr}{{\em Phys.\ Rev.\ }}
\newcommand{\cmp}{{\em Commun.\ Math.\ Phys.\ }}
\newcommand{\pl}{{\em Phys.\ Lett.\ }}
%
\bibitem{AbadMariusMehta}
C.A.~Abad and M.~Crainic, ``Representations up to homotopy of Lie
algebroids'', arXiv:0901.0319v1 [math.DG];\\
R.A.~Mehta, ``Q-algebroids and their cohomology'',
arXiv:math/0703234v1 [math.DG].
%\cite{Alexandrov:1995kv}
\bibitem{Alexandrov:1995kv}
  M.~Alexandrov, M.~Kontsevich, A.~Schwartz and O.~Zaboronsky,
  ``The Geometry of the master equation and topological quantum field theory,''
  Int.\ J.\ Mod.\ Phys.\ A {\bf 12} (1997) 1405
  [arXiv:hep-th/9502010].
  %%CITATION = HEP-TH 9502010;%%
  %
  %\cite{Axelrod:1991vq}
%
\bibitem{Axelrod:1989xt}
  S.~Axelrod, S.~Della Pietra and E.~Witten,
  ``Geometric quantization of Chern-Simons gauge theory,''
  J.\ Diff.\ Geom.\  {\bf 33}, 787 (1991).
  %%CITATION = JDGEA,33,787;%%
\bibitem{Axelrod:1991vq}
  S.~Axelrod and I.~M.~Singer,
  ``Chern-Simons perturbation theory,''
  Proceedings of the XXth International Conference on Differential Geometric Methods in Theoretical Physics, Vol. 1, 2 (New York, 1991), 3--45, World Sci. Publ., River Edge, NJ, 1992.
  [arXiv:hep-th/9110056].
  %%CITATION = HEP-TH/9110056;%%
   %
%\cite{Cattaneo:2001ys}
\bibitem{BN1}
D.~Bar-Natan,  ``On the Vassiliev Knot Invariants." Topology 34,
423-472, 1995.
%
\bibitem{wheeling}
D.~Bar-Natan, T.T.Q.~Le and D.P.~Thurston, ``Two applications of
elementary knot theory to Lie algebras and Vassiliev invariants'',
Geom. Topol. 7(2003) 1-31, arXiv:0204311v3 [math.QA].
%
\bibitem{BottTaubes}
R. Bott, C. Taubes, ``On the self-linking of knots," J. Math. Phys.
35, 5247-5287 (1994).
%\bibitem{Fuks}
% D.~B.~Fuks,
% ''Stable cohomologies of a Lie algebra of formal vector
%fields with tensor coefficients,''   Funktsional. Anal. i Prilozhen. {\bf 17} (1983), no. 4, 62.
%\cite{Cattaneo:1996pz}
\bibitem{Cattaneo:1996pz}
  A.~S.~Cattaneo,
  ``Cabled Wilson Loops in BF Theories,''
  J.\ Math.\ Phys.\  {\bf 37}, 3684 (1996)
  [arXiv:q-alg/9602015].
  %%CITATION = JMAPA,37,3684;%%
%
\bibitem{cattaneo-2002-2}
  A.~S.~Cattaneo, P.~Cotta-Ramusino and R.~Longoni,
  ``Configuration spaces and Vassiliev classes in any dimension,"
  Algebr.~Geom.~Topol.
   volume 2, 949, arXiv.org:math/9910139.
  %\cite{Cattaneo:2001ys}
\bibitem{Cattaneo:2001ys}
  A.~S.~Cattaneo and G.~Felder,
  ``On the AKSZ formulation of the Poisson sigma model,''
  Lett.\ Math.\ Phys.\  {\bf 56}, 163 (2001)
  [arXiv:math/0102108].
%%CITATION = LMPHD,56,163;%%
%\cite{Cattaneo:2009zx}
\bibitem{Cattaneo:2009zx}
  A.~S.~Cattaneo, J.~Qiu and M.~Zabzine,
  ``2D and 3D topological field theories for generalized complex geometry,''~Adv.Theor.Math.Phys.14:695-725,2010,~
  arXiv:0911.0993 [hep-th].
  %%CITATION = ARXIV:0911.0993;%%
%
\bibitem{ConantVogtmann}
J.~Conant and K.~Vogtmann, ``On a Theorem of Kontsevich,'' Algebr.
Geom. Topol. 3 (2003) 1167-1224, arXiv:math/0208169.
%
%\cite{Freed:1991wd}
\bibitem{Freed:1991wd}
  D.~S.~Freed and R.~E.~Gompf,
  ``Computer calculation of Witten's three manifold invariant,''
  Commun.\ Math.\ Phys.\  {\bf 141}, 79 (1991).
  %%CITATION = CMPHA,141,79;%%
%
\bibitem{hyperkahler}
A.~Fujiki, ``On primitively symplectic compact K\"ahler
V-manifolds of dimension four, Classification of algebraic and
analytic manifolds'' (Katata, 1982), pp. 71–250. Progr. Math.,
vol. 39.
Birkh\"auser Boston, Boston, MA (1983);\\
A.~Beauville, ``Vari\'et\'es k\"al\'eriennes dont la premi\`ere
classe de Chern est nulle'' J. Differ. Geom. 18, 755-782 (1983).
%\bibitem{FuksI}
% D.B.Fuks, ''Cohomology of infinite-dimensional Lie algebras and
% characteristic classes of foliations,'' Journal of Mathematical
% Sciences, Vol 11, {\bf 6}  (1979).
%
\bibitem{Hamilton-2005}
A.~Hamilton, ``A super-analogue of Kontsevich's theorem on graph
homology'', arXiv:math/0510390v1.
%
%\bibitem{hamilton-2006}
% A.~Hamilton and A.~Lazarev, ``Characteristic classes of  $A\sb\infty$
% algebras,''  J. Homotopy Relat. Struct. {\bf 3} (2008), no. 1, 65
%  [arXiv:math/0608395].
 %
\bibitem{Hamilton:2007tg}
  A.~Hamilton and A.~Lazarev,
  ``Graph cohomology classes in the Batalin-Vilkovisky formalism'',
  J.\ Geom.\ Phys.\  {\bf 59}, 555 (2009)
  [arXiv:math/0701825].
  %%CITATION = JGPHE,59,555;%%
%
%\cite{Ikeda:2010vz}
\bibitem{Ikeda:2010vz}
  N.~Ikeda and K.~Uchino,
  ``QP-Structures with Degree 4 and 4D Topological Field Theory,''
  arXiv:1004.0601 [hep-th].
  %%CITATION = ARXIV:1004.0601;%%
%
\bibitem{Jeffrey}
L.~Jeffrey, ``Chern-Simons-Witten Invariants of Lens Spaces and
Torus Bundles, and the Semiclassical Approximation," Commun. Math.
Phys. 147, 563-604 (1992).
%
 \bibitem{Jones}
 V. F. R. Jones, ``A polynomial invariant for knots via von Neumann
algebras," Bull. Amer. Math. Soc. (N.S.) Volume 12, Number 1
(1985), 103-111.
%
\bibitem{kapranov-1997}
 M.~Kapranov, ``Rozansky-Witten invariants via Atiyah classes,''
  Compositio Math. {\bf 115} (1999), no. 1, 71--113.
 [arXiv:alg-geom/9704009].
%
\bibitem{Kapustin:2009cd}
  A.~Kapustin and N.~Saulina,
  ``Chern-Simons-Rozansky-Witten topological field theory,''
  Nucl.\ Phys.\  B {\bf 823}, 403 (2009)
  [arXiv:0904.1447 [hep-th]].
  %%CITATION = NUPHA,B823,403;%%
\bibitem{Kontsevich:knot}
M.~Kontsevich,  ``Vassiliev's Knot Invariants." Adv. Soviet Math.
16, Part 2, pp. 137-150, 1993.
%
\bibitem{Kontsevich:formal}
  M.~Kontsevich,
  ``Formal (non)-commutative symplectic geometry,'' The
  Gelfand Mathematical Seminars, 1990 - 1992, Birkh\"auser (1993),
  173 - 187.
  %
\bibitem{Kontsevich:Feynamn}
  M. Kontsevich, ``Feynman diagrams and low-dimensional topology,''
  First European Congress of Mathematics, 1992, Paris, Volume II,
  Progress in Mathematics 120, Birkh\"auser 1994, 97 - 121.
%
% \bibitem{kontsevich1}
%M.~Kontsevich,
% ``Rozansky-Witten invariants via formal geometry,"
% Compositio \ Math.\  {\bf 115} (1999) 115
% [arXiv:dg-ga/9704009].
%  %
%  \bibitem{lyakhovich-2009}
%  S.~L.~Lyakhovich, E.~A.~Mosman and A.~A.~Sharapov,
%  ``Characteristic classes of Q-manifolds: classification and applications,''
%  arXiv:0906.0466 [math-ph].
%  %%CITATION = ARXIV:0906.0466;%%
%
%  \bibitem{kontsevich2}
 % Kontsevich, M. "Vassiliev's Knot Invariants." Adv. Soviet Math. 16, Part 2, pp. 137-150, 1993
%
%  \bibitem{morita}
%  S.~Morita,
%  ``Geometry of characteristic classes,''
%   {\it Iwanami Series in Modern Mathematics}, American Mathematical Society,
%    Providence, RI, 2001. xiv+185 pp.
%    %
%  %\cite{Qiu:2009zv}
%\cite{Labastida:1998ud}
\bibitem{Labastida:1998ud}
  J.~M.~F.~Labastida,
  ``Chern-Simons gauge theory: Ten years after,''
  arXiv:hep-th/9905057.
  %%CITATION = HEP-TH/9905057;%%
\bibitem{AKSZ_RW}
  J.~Qiu and M.~Zabzine,
  ``On the AKSZ formulation of the Rozansky-Witten theory and beyond,''
  JHEP {\bf 0909}, 024 (2009)
  [arXiv:0906.3167 [hep-th]].
  %%CITATION = JHEPA,0909,024;%%
%
\bibitem{QiuZabzine:2009rf}
  J.~Qiu and M.~Zabzine,
  ``Odd Chern-Simons Theory, Lie Algebra Cohomology and Characteristic
  Classes,''~Commun.Math.Phys.300:789-833,2010,
  arXiv:0912.1243 [hep-th].
  %%CITATION = ARXIV:0912.1243;%%
%
\bibitem{roberts-2006}
  J.~Roberts and S.~Willerton,
  ``On the Rozansky-Witten weight systems,"
  arXiv.org:math/0602653.
\bibitem{roberts-2001}
  J.~Roberts, ``Rozansky-Witten theory,"
  arXiv.org:math/0112209.
%
%
%\cite{Roytenberg:2002nu}
%\bibitem{Roytenberg:2002nu}
%  D.~Roytenberg,
%  ``On the structure of graded symplectic supermanifolds and Courant
%  algebroids,''   in: \textit{Quantization, Poisson Brackets and Beyond}, Theodore Voronov (ed.),
%   \textit{Contemp. Math}, Vol. 315, Amer. Math. Soc., Providence, RI, 2002,
%  [arXiv:math/0203110].
%  %%CITATION = MATH/0203110;%%
%
% \cite{Roytenberg:2006qz}
\bibitem{Roytenberg:2006qz}
  D.~Roytenberg,
  ``AKSZ-BV formalism and Courant algebroid-induced topological field
  theories,''
  Lett.\ Math.\ Phys.\  {\bf 79} (2007) 143
  [arXiv:hep-th/0608150].
%  %%CITATION = LMPHD,79,143;%%
  %
%\cite{Rozansky:1994ba}
\bibitem{Rozansky:1994ba}
  L.~Rozansky,
  ``A Contribution of the trivial connection to the Jones polynomial and
  Witten's invariant of 3-d manifolds. 2,''
  Commun.\ Math.\ Phys.\  {\bf 175}, 297 (1996)
  [arXiv:hep-th/9403021].
  %%CITATION = CMPHA,175,297;%%
%
\bibitem{Rozansky:1996bq}
  L.~Rozansky and E.~Witten,
  ``Hyper-K\"ahler geometry and invariants of three-manifolds,''
  Selecta Math.\  {\bf 3} (1997) 401
  [arXiv:hep-th/9612216].
  %%CITATION = SMATF,3,401;%%
  %
  \bibitem{sawon}
J.~Sawon, ``Rozansky-Witten invariants of hyperk\"ahler
manifolds,''
 PhD thesis, Oxford 1999.
%
\bibitem{Schwarz:1992nx}
  A.~S.~Schwarz,
  ``Geometry of Batalin-Vilkovisky quantization,''
  Commun.\ Math.\ Phys.\  {\bf 155}, 249 (1993)
  [arXiv:hep-th/9205088].
  %
%\cite{Schwarz:1999vn}
\bibitem{Schwarz:1999vn}
  A.~S.~Schwarz,
  ``Quantum observables, Lie algebra homology and TQFT,''
  Lett.\ Math.\ Phys.\  {\bf 49}, 115 (1999)
  [arXiv:hep-th/9904168].
  %%CITATION = LMPHD,49,115;%%
%
%\cite{Thompson:2000pw}
\bibitem{Thompson:2000pw}
  G.~Thompson,
  ``Holomorphic vector bundles, knots and the Rozansky-Witten invariants,''
  Adv.\ Theor.\ Math.\ Phys.\  {\bf 5}, 457 (2002)
  [arXiv:hep-th/0002168].
  %%CITATION = 00203,5,457;%%
\bibitem{Thurston}
 D.~P.~Thurston, ``Integral Expressions for the Vassiliev Knot
 Invariants'', arXiv:math/9901110.
%
%
\bibitem{Vass}
V.A.Vassiliev, ``Complements of Discriminants of Smooth Maps:
Topology and Applications," Translations of Mathematical Monographs
Vol 98, AMS Providence, Rhode Island.
%
\bibitem{Vogel}
P.~Vogel, ``Algebraic Structures on Modules of Diagrams'',
preprint 1997.
%\bibitem{Voronov:2001qf}
%  T.~Voronov,
%  ``Graded manifolds and Drinfeld doubles for Lie bialgebroids,''
%  in: \textit{Quantization, Poisson Brackets and Beyond}, Theodore Voronov (ed.), \textit{Contemp. Math}, Vol. 315, Amer. Math. Soc., Providence, RI, 2002,
%[arXiv:math/0105237].
%  %%CITATION = MATH/0105237;%%
%%
\bibitem{Witten:1988hf}
  E.~Witten,
  ``Quantum field theory and the Jones polynomial,''
  Commun.\ Math.\ Phys.\  {\bf 121}, 351 (1989).
  %%CITATION = CMPHA,121,351;%%
\end{thebibliography}
\end{document}